# Real-time Monitoring of Economic Shocks using Company Websites

Michael König[1,2,3], Jakob Rauch[2*], Martin Wörter[1]

[1]ETH Zurich, Swiss Economic Institute (KOF), Leonhardstrasse 21, Zurich, 8092, Switzerland.
[2]Tinbergen Institute and Department of Spatial Economics, Vrije Universiteit Amsterdam, De Boelelaan 1105, Amsterdam, 1081 HV, The Netherlands.
[3]Centre for Economic Policy Research (CEPR), Great Sutton St. 33, London, EC1V 0DX, United Kingdom.

*Corresponding author(s). E-mail(s): j.m.rauch@vu.nl;
Contributing authors: koenig@kof.ethz.ch; woerter@kof.ethz.ch;

**Abstract**

Understanding the effects of economic shocks on firms is critical for analyzing economic growth and resilience. We introduce a Web-Based Affectedness Indicator (WAI), a general-purpose tool for real-time monitoring of economic disruptions across diverse contexts. By leveraging Large Language Model (LLM) assisted classification and information extraction on texts from over five million company websites, WAI quantifies the degree and nature of firms' responses to external shocks. Using the COVID-19 pandemic as a specific application, we show that WAI is highly correlated with pandemic containment measures and reliably predicts firm performance. Unlike traditional data sources, WAI provides timely firm-level information across industries and geographies worldwide that would otherwise be unavailable due to institutional and data availability constraints. This methodology offers significant potential for monitoring and mitigating the impact of technological, political, financial, health or environmental crises, and represents a transformative tool for adaptive policy-making and economic resilience.

**Keywords:** large language models, natural language processing, crisis, economic shocks, economic monitoring, Covid-19

Economic shocks, whether driven by public health crises, technological disruptions, geopolitical conflicts, or climate events, pose significant challenges to businesses and policymakers alike. Timely and accurate monitoring of these shocks is critical for crafting effective responses and enhancing economic resilience. However, traditional methods for measuring the impacts of such disruptions – such as surveys and administrative data – are often limited by costs, time lags, and coverage.

In this study, we introduce the *Web-Based Affectedness Indicator* (WAI), a scalable and cost-effective tool for real-time monitoring of economic disruptions at the firm level. By analyzing textual data from millions of company websites, WAI provides granular insights into how firms experience and respond to external shocks. This methodology overcomes traditional limitations by leveraging ubiquitous online content and state-of-the-art natural language processing (NLP) models to generate a dynamic and comprehensive view of economic affectedness.

While the Covid-19 pandemic[1–3] serves as a prominent example to illustrate WAI's capabilities, the method's applicability extends far beyond public health crises. WAI can provide information on a wide range of challenges, including supply chain disruptions, financial crises, and climate-related shocks. By offering real-time, high-resolution insights, it provides a transformative tool for policymakers, businesses, and researchers seeking to understand and mitigate the impacts of global disruptions.

Our novel method substantially improves on existing approaches in the literature. Several surveys have been conducted to analyze how firms have been affected by the Covid-19 pandemic.[4–9] In addition, there exist several survey-based studies investigating the relationship between productivity and firm resilience during the Covid-19 crisis.[10–13] As an alternative to survey data, administrative or accounting data can be used to analyze the economic impact of the Covid-19 crisis.[14,15] Various studies further analyze the effectiveness of Covid-19 policy support measures.[16,17] The above mentioned data sources are undeniably valuable. However, they come with some limitations. Surveys, for example, can be expensive to conduct and can suffer from low participation rates, particularly during periods of a crisis. Accounting data, at the firm or national level, might be difficult to access, have only partial coverage, and typically becomes available only with a significant delay. Our study aims at overcoming these limitations and adds to an emerging body of literature that uses "text as data"[18,19] to examine the economic repercussions of the Covid-19 crisis, while broadening the existing literature that has been limited to a single country or publicly listed firms.[20–26] Notably, sample sizes in the above mentioned studies typically range in a few thousand companies, while we cover over 5 million public and private companies.

The WAI workflow is illustrated in Figure 1 (see Methods for further details). Our content extraction process involves downloading and analyzing company websites from the internet archive CommonCrawl, and identifying paragraphs mentioning Covid-related keywords.[1] Using a large language model (Llama 3.1) through a simple few-shot prompt method,[27] we then classify the severity of impact on firms by scoring the affectedness and tagging specific aspects such as facility closures or supply chain issues. By analyzing multiple crawls of the same company's website in combination with location and industry information, we are able to track changes over time and analyze firm-level impacts across sectors and space.

While almost all firms maintain a website nowadays, it is not obvious that the content on a website can be used as an economic shock indicator. To validate WAI as a real-time economic shock monitoring tool, we compare it with (i) alternative measures for Covid-19 policy interventions that are only available at the aggregate (state or country) level and (ii) firm performance measures from balance sheet data (sales growth or stock market valuations) for a subset of publicly listed firms.[2]

For our validation exercise, we first consider Covid-19 pandemic policy responses in the US and Europe. Figure 2 shows the economic affectedness from Covid-19 across the US and the four largest states within the US (California, Florida, Texas and New York). Figure 3 shows the affectedness in Europe and among major European economies (United Kingdom, Germany, France and Spain). Both figures show the share of companies with the highest level of impact according to WAI (a score of 3 on a scale from 0 to 3) for each country and crawl. As an alternative affectedness measure, that is available at this level of aggregation (but not at the firm-level), the figures also show the government response stringency index[3] of the Oxford Covid-19 Government Response Tracker (OxCGRT).[28] Moreover, the number of analyzed firms, and the Pearson correlation coefficient between WAI and OxCGRT (of the first differences of the two times series) are shown. The figures illustrate a strong correlation between the two measures. This shows that when aggregating WAI at the state level it can accurately measure the economic shock stemming from Covid-19 policy interventions.[4]

---

[1] See SI Appendix A.2 for a complete list of keywords and languages considered.

[2] We also compared WAI with the proportion of Covid-related keywords in all the words of the "risk assessments" in the business reports (10-K reports) of publicly listed firms in the US,[21] and find a significant positive correlation (with a correlation coefficient of 0.427).

[3] The stringency index aggregates information from school closures, workplace closures, cancellation of public events, restrictions on public gatherings, closures of public transport, stay-at-home requirements, public information campaigns, restrictions on internal movements, and international travel controls. A higher score indicates a stricter response, with a minimum of 0 and a maximum of 100.

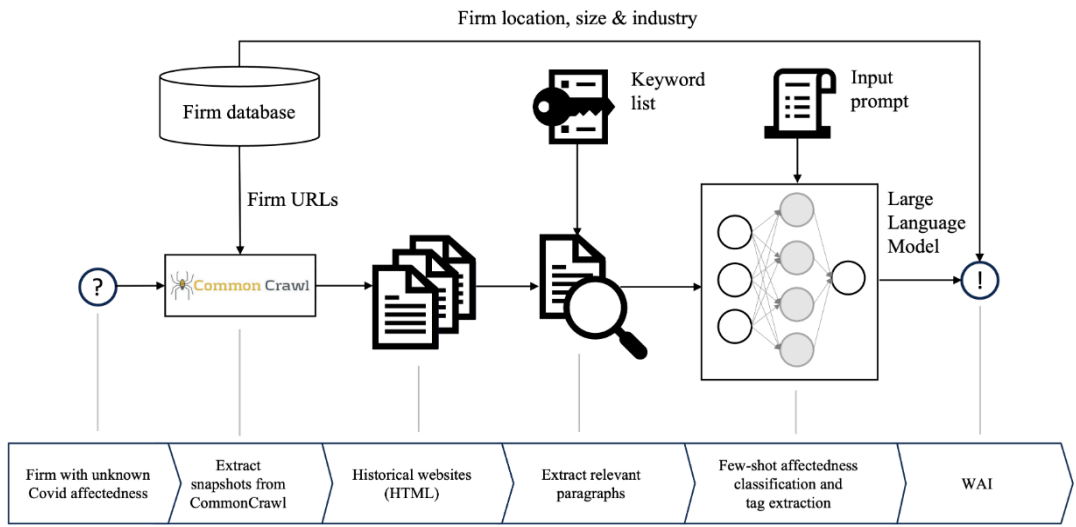

**Figure 1**: Illustration of the WAI workflow (see Methods). We extract content from company websites in CommonCrawl, classify Covid-19 impact using a large language model, and track changes over time to analyze firm-level impacts across sectors and different geographies.

Comparing the OxCGRT stringency index with WAI in Figures 2 and 3, respectively, we see that in the early stages of the pandemic, the stringency index shows a steeper increase than WAI and that WAI starts to increase only with a small time lag after the stringency index. This reflects a delay in the companies' affectedness from the policy interventions possibly due to the existence of inventory stocks or excess capacities that allowed firms to absorb temporarily the impact of the shock. Moreover, disruptions in the companies' operations might only surface with a delay, or firms might initially have anticipated that the polices would only be in place for a short time period. However, as policy interventions persisted over time, WAI and stringency index become more closely aligned. In particular, as more drastic measures such as workplace closures and stay-at-home requirements were lifted, WAI reacts almost instantaneously by showing lower levels of affectedness.

To further evaluate the relevance of WAI for individual firm performance we run a panel regression with quarterly sales growth (in %) and stock market valuations as dependent variables using WAI intensity levels of Covid-19 impact – ranging from mild, to moderate and severely affected[5] – as predictors with various controls (see Methods). The estimation results are shown in Figure 4 for the sample periods 2017 to 2022 with quarterly balance sheet data for around 30 thousand publicly listed firms (from S&P's Compustat North America and Global). In the simplest specification, we include firm-level fixed effects, which capture factors such as a firm's propensity to use its website to communicate its exposure to external shocks, and quarterly time fixed effects, which capture firm-invariant time trends such as the cross-regional temporal evolution of the pandemic. The estimated coefficients shown in the left panel in Figure 4 can be interpreted as follows: Companies that were only slightly affected by Covid-19 did not suffer any significant sales growth losses. By contrast, moderately affected companies recorded a significant 12.1% decrease in sales growth (annualized) compared to not or positively affected companies. Severely affected companies show a slightly higher sales growth loss of around 15%. In an extended specification we include additional controls for government policies as well as the epidemiological development measured by the number of people who died from Covid-19 in a country per month. Finally, we include country-industry-quarter-specific fixed effects. This does not only absorb all additional controls in the previous specification but also unobserved heterogeneity in policies and the development of the pandemic in each country. This leads to slightly smaller marginal effects. Slightly affected firms have 7.8% lower annualized sales growth than unaffected or positively affected companies and severely affected firms experienced 11.7% lower sales growth in this most stringent estimate.

---

[4] A complete overview of all US states and other countries considered in our data sample can be found in SI Appendices B.1 and B.2, respectively.

[5] The group of unaffected or positively affected companies serves as the reference category.

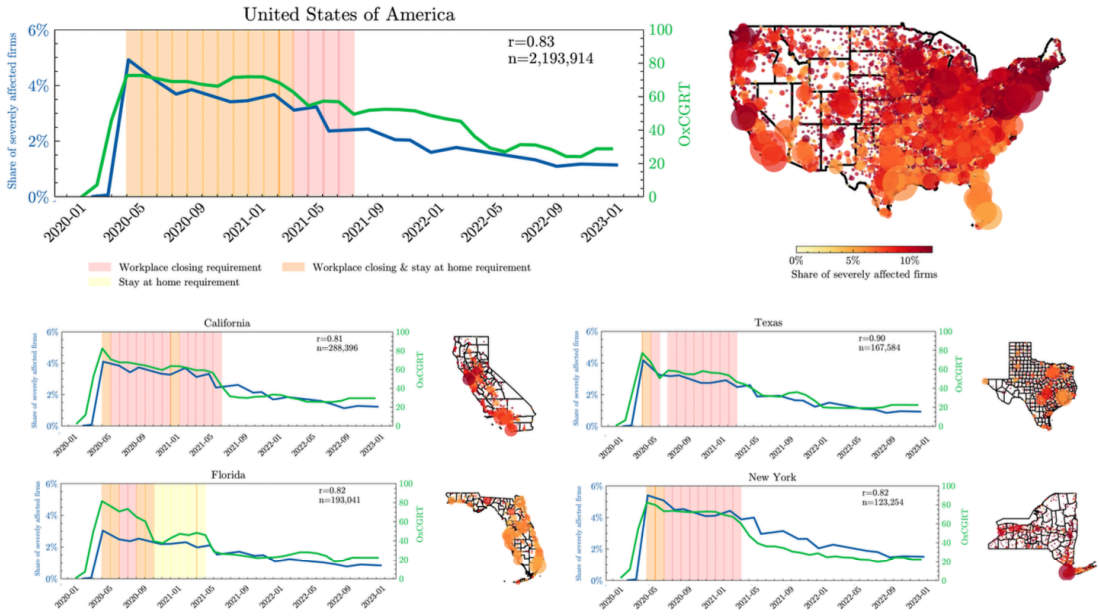

**Figure 2**: Left panels show the WAI (share of severely affected firms) and OxCGRT government response stringency index in the US and the four largest states within the US (California, Florida, Texas and New York) together with the number of analyzed firms ($n$), and the Pearson correlation coefficient ($r$) between the first differences (over time) of WAI and OxCGRT. Time periods in which workplace closing or stay at home requirements were implemented are indicated with vertical bars. Right, spatial maps indicating WAI share of severely affected firms at the city level, with the size of each city proportional to the count of analyzed firms in it.

In the estimates for stock returns (shown in the right panel in Figure 4), we see in the strictest estimation model (with full fixed effects) that even slightly affected companies show significantly lower stock returns of annualized 3% compared to companies that are not affected or are positively affected. As expected, moderately affected companies and severely affected companies show even stronger losses. They amount to 3.3% for moderately affected companies and 4.4% for severely affected companies.[6]

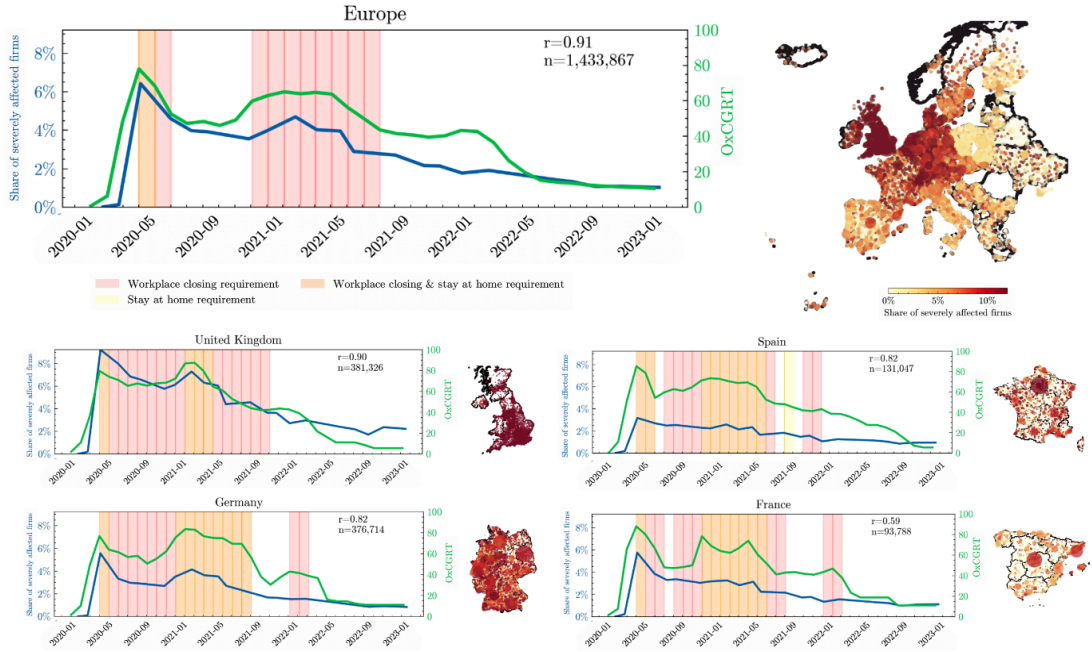

**Figure 3**: Covid-19 affectedness based on WAI and OxCGRT government response stringency index in Europe and four European economies (United Kingdom, Germany, France and Spain). OxCGRT stringency for Europe was constructed as an (unweighted) average of the stringency in all European countries.

The above discussion demonstrates that WAI accurately predicts a firm's sales growth and stock returns across all specifications. This confirms that WAI delivers timely, firm-level insights into exposure and its associated performance effects – insights that go beyond what can be inferred from monitoring government policies or epidemiological trends alone. The indicator is also precise enough to distinguish the effects even in terms of the severity of affectedness. Our results thus align with previous studies showing that returns on assets[14] or aggregate GDP growth[15] deteriorated during the Covid-19 pandemic.

Additional robustness checks for different LLMs, time horizons, sectors and specifications can be found in SI Appendix C.2. In SI Appendix C.2.1 we use an alternative LLM (ChatGPT) to compute WAIs and find that the estimates show a high degree of similarity. In SI Appendix C.2.2 we estimate the model for different time periods and find that the coefficients for the WAIs tend to be higher but less precisely measured when we consider a shorter sample period from 2018 to 2021, or from 2019 to 2020. In SI Appendix C.2.3 we estimate the model for the manufacturing and services sectors separately. We find that the estimated coefficients for sales growth are higher in the manufacturing sector than in the services sector. This result is consistent with the fact that the manufacturing sector is particularly vulnerable to supply chain shocks due to its reliance on complex, global supply chains and just-in-time inventory systems, and that supply chains being identified as the most significant issue mentioned by firms on their websites (see Figure 5). Finally, in SI Appendix C.2.4 we estimate a model that includes a one-period lag of the dependent variable to account for potential serial correlation. We find that our baseline estimates remain robust under this alternative specification.

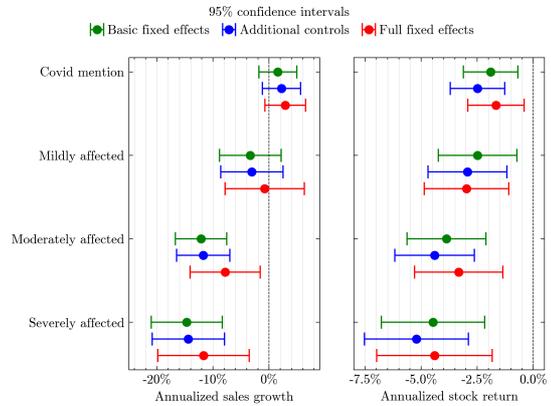

**Figure 4**: Estimation results for sales growth and stock returns over the periods 2017-2022 (see Eqs. (1) to (3) in Methods, and SI Appendix C.1 for further details). The full set of estimation results can be found in SI Appendix Table C.1.

---

[6]The detailed estimation results for quarterly sales growth and quarterly stock returns can be found in Table C.1 in the supplementary information (SI) Appendix C.1.

After having provided evidence for the validity of WAI we can now analyze its output more closely. First, we can use the LLM to assign tags indicating the type of problems firms were experiencing during the Covid-19 pandemic. These tags are shown in Figure 5 across countries. The three most prominent tags were related to supply chain issues followed by closure and hygiene measures (such as quarantine or social distance; see Methods). Moreover, up to 60% of the firms reported supply chain problems in the US, Canada and the UK. This resonates with results from recent surveys among small-to-medium sized businesses,[29] but covers a much broader range of firms.

Second, we can analyze which sectors were affected the most from the Covid-19 pandemic. Figure 6 shows a ranking of the most affected sectors, based on the WAI share of severely affected firms by industry, where each firm is weighted by its number of employees. This measures can be interpreted as the share of employees working in severely affected firms in each industry. We find that the most affected sectors were 'Accommodation' (42%), followed by 'Arts, Entertainment and Recreation and Human Health' (40%), and 'Social

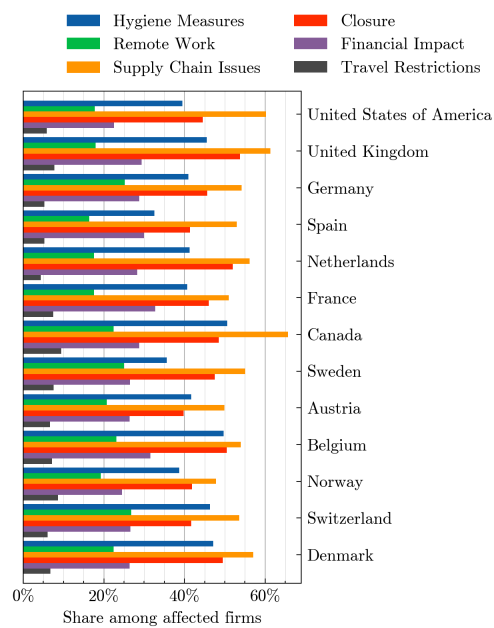

**Figure 5**: Tags assigned by the LLM for how firms were affected, aggregated by country (see Methods).

Work Activities' (39%). The least affected sectors were 'Energy Utilities (19%), 'Wholesale Trade' (10%) and 'Construction' (7%). This is broadly in line with previous studies based on risk assessments of publicly listed firms in the US,[20,21] but our analysis covers not only public but also private firms.

This paper presents a novel method for assessing the economic impact of social and economic shocks on firms using web-based data. By analyzing over five million company websites, we develop WAI and use it to measure the firm-level impact of the Covid-19 pandemic. The reliability of WAI is demonstrated through its strong correlation with pandemic containment measures and its ability to accurately capture the impact on business performance.

Unlike traditional data collection methods, our web-based indicator is not only comprehensive and reliable but provides also real-time information, as many companies immediately report the impact of shocks on their websites to inform their customers or other relevant parties. Real-time information shortens the trial-and-error phase of policy formulation, improving the accuracy of responses and safeguarding public budgets.

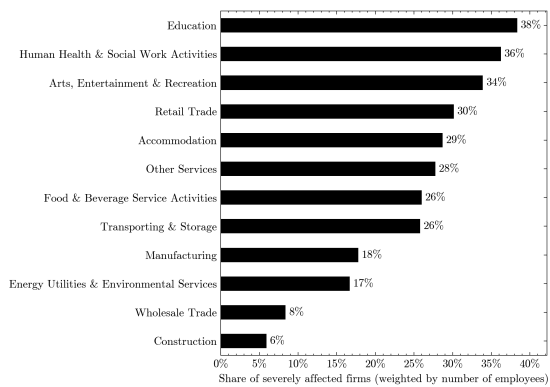

**Figure 6**: The share of severely affected firms by industry (see Methods).

A cross-country comparison suggests that WAI performs most effectively in countries with widespread internet usage and a predominantly English-speaking population, such as the US, Canada, the UK, and Ireland (cf. SI Appendix B.2). This limitation could be attributed to the fact that the LLMs employed in this study are primarily trained on English-language corpora, which may give them an advantage in these countries. However, with an increasing digitization across countries world-wide and improvements in multi-lingual LLMs we expect these differences to become less prominent over time, and thus making WAI an even more valuable tool in the future.

The findings of this study mark a significant advancement in real-time economic monitoring at the company level, offering valuable insights for policy-making. This capability is essential for enhancing the resilience of economies to future global shocks, whether they are health-related, technological, political, financial, or climate-driven (e.g., "Disease X"[30], tariffs, escalating trade wars, etc.).

# Methods

### Data collection and processing

We use multiple data sources and different methods to investigate whether and how companies were affected by the Covid-19 pandemic. We use information from over five million company websites worldwide to determine whether the pandemic caused economic disruptions at the firm-level. To measure the impact of the pandemic on firm performance, we use sales information from S&P's Compustat North America and Global financials databases. Information on government interventions is taken from the Oxford Covid-19 Government Response Tracker (OxCGRT).[28] In the following, we describe these datasets in more detail, demonstrate how we analyze company websites and what estimation techniques are used to relate the development of company performance to the Covid-19 exposure.

**Historical company website information.** We use the CommonCrawl dataset to access historical information from company websites. CommonCrawl is an extensive and constantly updated collection of web data that covers a large part of the web content and allows access to the historical content of a website. This makes it possible to follow the development of website content over time. CommonCrawl is a 501(c)(3) nonprofit and all its data is open source.[7] The data is organized into crawl snapshots each containing around 30 million domains and 3 billion URLs.[8] The crawls are done at a close-to-monthly frequency. For the period of concern, January 2020 to December 2022, the available crawls are listed in Extended Data Table 1.

Information about the crawling process, selection of seed URLs, crawling frequency, and politeness policies[9] can be found on the CommonCrawl website. We built a tool to download

---

[7] https://github.com/commoncrawl

[8] https://commoncrawl.github.io/cc-crawl-statistics

[9] In particular the crawling respects the `robot.txt` of the accessed websites.

and process the historical website information. The tool is published open source on a GitHub website.[10]

We start our analysis with a list of company website domains from Bureau Van Dijk's Orbis database, which is a comprehensive and widely used source of public and private company data.[31] We restrict our selection to companies in the database with a website address, location information, and at least 5 employees. This yields 8,164,172 unique domains.[11] Of these, 5,592,178 can be found in CommonCrawl.[12] A website not being indexed by CommonCrawl can have multiple reasons, such as the website being unreachable/unmaintained, the administrator not allowing crawls of the website, or the website simply not being linked to any other part of the internet so that the crawler cannot find it. Note that domains that are part of a crawl in e.g. January are not always part of the next crawl in February. However, the domain overlap between crawls – as measured by the Jaccard similarity shown in Extended Data Figure 7 – is high, indicating that we can follow the same firms over multiple observation periods. We also construct a measure for the content overlap of a firm's website between crawls. CommonCrawl assigns unique so-called 'content digest' IDs to identify (exactly) duplicate captures of a website. For each firm, we calculate a 'content digest heartbeat', which is the share of crawls between which the content on the firm's website has changed. For instance, a value of 0.75 means that the content has changed between 18 out of the 24 total analyzed crawls. On average, firms change their website content often: the median firm has a heartbeat of 0.97 (the full distribution is shown in Extended Data Figure 8).

For each of the domains we have information for in CommonCrawl, we download all subpages (such as `apple.com/home`, `apple.com/newsroom`) available, up to a total of 50. If there are more than 50 subpages available, we select the 50 shortest URLs - these subpages will be higher up in the sitemap and therefore contain more generally relevant information. In addition, we select *all* URLs that contain one of the Covid-19 keywords in any language (such as `aholddelhaize.com/covid-19`). In most cases, we find relevant information on either subpages with keyword-containing URLs, the landing page, or a '/news' or similarly titled subpage. Therefore, we would not expect significant performance benefits from increasing the number of analyzed shortest subpages beyond 50. In the next step, we parse the HTML code of each website and extract paragraphs containing at least one of a list of Covid-related keywords, translated into 65 languages.[13] The paragraphs are then processed using Llama 3.1, an open-source large language model (LLM) developed by Meta AI. At the time of this writing, the model is at the accuracy-efficiency frontier among open-source LLMs.[14] To speed up processing, we use 8-bit floating point quantization which leads only to a minimal loss of quality. We use a custom logits processor that forces the model to return outputs in proper JSON format along with the following parameters: a temperature setting of 0 (because we want the model to choose the likeliest outputs in terms of affectedness and tags and not be creative), 64 maximum output tokens (which is more than sufficient to assign multiple tags if required), a fixed seed and stop characters '0' and '}', meaning the LLM will stop generating after either affectedness is detected as 0 or reaching the end of proper JSON output. We use the few-shot prompt shown in Extended Data Figure 9,[15] in which we insert the Covid-mentioning paragraph as '<Input paragraph>'. The LLM returns a number between 0 and 3 indicating the severity of affectedness, as well as a list of tags indicating how the firm was affected. For most firms that do mention Covid, there are multiple text passages (often also from different subpages) that contain Covid keywords.

---

[10] https://github.com/jakob-ra/cc-download

[11] An overview of the countries covered can be found in SI Appendix A.1.

[12] Note that for some companies we have multiple websites: The 5,592,178 unique domains correspond to 5,429,830 unique firms. In the case of multiple websites per firm, we take that firms affectedness indicator to be the maximum of across its websites.

[13] This was done using the following prompt for the large language model GPT-3.5-turbo: Translate the following keywords into: 'corona', 'covid', 'covid-19', 'sars-cov-2', 'coronavirus', 'pandemic' Please only output the translated keywords, each on a new line. The resulting keywords can be found in SI Appendix A.2.

[14] For comparison, we also repeat the analysis with OpenAI's closed-source GPT-4o mini model and find that our approach matches or exceeds its quality. The results can be found in SI Appendix Table C.2.

[15] Few shot prompting has been shown to lead to comparable performance to fine-tuning,[27] especially for less complex information extraction tasks like the one considered here.[32]

The severity value for the firm for one time period is the maximum over all text passages mentioning Covid, and the tags are the union of tags assigned to all passages.

**Tag construction.** In the prompt we instructed the LLM to assign tags describing specific ways a firm was affected by Covid-19. We have summarized the tags identified by the LLM under the following umbrella terms: Supply chain issues, closure, remote work, hygiene measures, travel restrictions and financial impact. The individual tags proposed by the LLM that fall under these umbrella terms can be found in Extended Data Figure 10.

**Company performance information.** Our study encompasses data from S&P's Compustat North America database, which includes 43,963 publicly listed US firms, and the Compustat Global database, covering 14,076 publicly listed non-US firms, for the years 2017 to 2022, using consolidated accounts. We conducted an in-depth analysis of quarterly financials, focusing on sales, number of employees, and total assets. To measure the financial performance across quarters, we calculated quarter-to-quarter growth rates after converting all financial levels to USD, utilizing currency exchange rates from Yahoo Finance. Additionally, we examined stock returns on the last day of each quarter, making adjustments for stock splits and applying the total return factor. For a comprehensive overview, we categorized the data by city, country, and industry, converting NAICS data to the NACE classification using established correspondence tables.[16]

**Covid-19 policy measures.** Information on policy interventions in response to the Covid-19 pandemic is obtained from the Oxford Covid-19 Government Response Tracker (OxCGRT).[28]

*Econometric Analysis*

We collected historic websites for 42,644 out of 52,245 companies listed in Compustat North America and Global through CommonCrawl, covering 81.6% of the companies. This aligns with previous research findings on CommonCrawl's representativeness.[17]

The benchmark model used to analyze how our indicators track firm performance is expressed as:

$$\begin{aligned}\Delta \log y_{i,t} &= \log y_{i,t} - \log y_{i,t-1} \\ &= \alpha + \delta_i + \delta_t + Z'_{i,t-1}v + \beta \text{Covid mention}_{i,t} \\ &\quad + \gamma_1 \text{Mildly affected}_{i,t} + \gamma_2 \text{Moderately affected}_{i,t} \\ &\quad + \gamma_3 \text{Severely affected}_{i,t} + \epsilon_{i,t}\end{aligned} \quad (1)$$

where $y_{i,t}$ denotes either sales or stock price for firm $i$ in quarter $t$. The dependent variable $\Delta \log y_{i,t}$ is either log sales growth or log stock returns. The coefficient $\delta_i$ denotes a firm fixed effect, $\delta_t$ are quarter fixed effects, $Z_{i,t-1}$ are time-varying firm characteristics (log total assets), and $\epsilon_{i,t}$ is an error term. The dependent variable is transformed into a percentage so that the estimated coefficients can be more easily interpreted. 'Covid mention' is a dummy for whether a company mentions Covid-19 on its websites, and the affectedness severity (ranging from 'Mildly affected', 'Moderately affected' to 'Severely affected') stems from the LLM output.

We also estimate a model with additional controls for government policies and the development of the pandemic:

$$\begin{aligned}\Delta \log y_{i,t} &= \alpha + \delta_i + \delta_t + W'_{c,t}u + Z'_{i,t-1}v + \beta \text{Covid mention}_{i,t} \\ &\quad + \gamma_1 \text{Mildly affected}_{i,t} + \gamma_2 \text{Moderately affected}_{i,t} \\ &\quad + \gamma_3 \text{Severely affected}_{i,t} + \epsilon_{i,t}\end{aligned} \quad (2)$$

where $W_{c,t}$ are variables relating to government policies[18] and epidemiological variables in country $c$ and quarter $t$. Finally, we estimate a model with full fixed effects as follows:

$$\begin{aligned}\Delta \log y_{i,t} &= \alpha + \delta_i + \delta_{s,c,t} + Z'_{i,t-1}v + \beta \text{Covid mention}_{i,t} \\ &\quad + \gamma_1 \text{Mildly affected}_{i,t} + \gamma_2 \text{Moderately affected}_{i,t} \\ &\quad + \gamma_3 \text{Severely affected}_{i,t} + \epsilon_{i,t}\end{aligned} \quad (3)$$

where $\delta_{s,c,t}$ are sector-country-quarter (index by $s, c, t$, respectively) fixed effects, sector/industry being measured at the NACE 2-digit level. This model is useful to see whether our indicators are informative even beyond all country and industry specific time trends. This empirical approach allows us to determine whether the indicator is accurate enough to reflect, for example, the different degrees to which businesses within a sector (e.g. restaurants) are affected, given the state of the pandemic and sector-country-specific measures at a given point in time. The estimation results are shown in Figure 4. A complete overview of the estimation results across the different specifications in Eqs. (1), (2) and (3) can be found in SI Appendix C.1.

---

[16] https://www.census.gov/naics/concordances/2017_NAICS_to_ISIC_4.xlsx)

[17] For example, 80-85 percent of the Alexa top 1-million websites can be found in CommonCrawl. See: https://commoncrawl.github.io/cc-crawl-statistics/plots/tld/comparison.html.

[18] These are national measures aimed at preventing the spread of Covid-19, such as workplace closing and stay-at-home requirements, as well as policies aimed at supporting the economy through a fiscal response

**WAI variation by industry.** The industries in Figure 6 follow NACE sections, however we split up some sections into their constituent 2-digit codes: Accomodation (55), Food & Beverage Service Activities (56), Wholesale trade (except of motor vehicles and motorcycles, 46), Retail trade (except of motor vehicles and motorcycles, 47). We also combine the following NACE sections into 'Other Services': Financial and insurance activities, Real estate activities, Information and communication, Professional, scientific and technical activities, Administrative and support service activities, Other services activities (sections J through N plus S) and the following sections into 'Energy Utilities and Environmental Services': Electricity, gas, steam and air conditioning supply, and Water supply; sewerage; waste managment and remediation activities sections (D and E).

# Declarations

**Data availability.** The dataset generated and analyzed in the current study is publicly accessible via an interactive online dashboard available at https://covid-explorer.kof.ethz.ch/.

**Acknowledgments.** We would like to thank Elliot Ash, Gert Buiten, Marzio di Vece, Francois Lafond, Frank Takes, Tobias Reisch and Fernando Vega-Redondo for the helpful comments, and Oliver Müller for the assistance with the dashboard.

**Authors' contributions.** J.R. collected, verified, cleaned and merged data. All authors designed analysis, interpreted results, and wrote the paper.

**Competing interests.** The authors declare no competing interests.

**Ethics approval.** Not applicable.

**Code availability.** All codes for reducibility of the results presented in this paper are available at https://github.com/jakob-ra/firm-level-web-indicator

**Extended Data Table 1**: CommonCrawl dataset snapshots.

| Time Period | Snapshot ID |
| --- | --- |
| January 2020 | CC-MAIN-2020-05 |
| February 2020 | CC-MAIN-2020-10 |
| March/April 2020 | CC-MAIN-2020-16 |
| May/June 2020 | CC-MAIN-2020-24 |
| July 2020 | CC-MAIN-2020-29 |
| August 2020 | CC-MAIN-2020-34 |
| September 2020 | CC-MAIN-2020-40 |
| October 2020 | CC-MAIN-2020-45 |
| November/December 2020 | CC-MAIN-2020-50 |
| January 2021 | CC-MAIN-2021-04 |
| February/March 2021 | CC-MAIN-2021-10 |
| April 2021 | CC-MAIN-2021-17 |
| May 2021 | CC-MAIN-2021-21 |
| June 2021 | CC-MAIN-2021-25 |
| July/August 2021 | CC-MAIN-2021-31 |
| September 2021 | CC-MAIN-2021-39 |
| October 2021 | CC-MAIN-2021-43 |
| November/December 2021 | CC-MAIN-2021-49 |
| January 2022 | CC-MAIN-2022-05 |
| May 2022 | CC-MAIN-2022-21 |
| June/July 2022 | CC-MAIN-2022-27 |
| August 2022 | CC-MAIN-2022-33 |
| September/October 2022 | CC-MAIN-2022-40 |
| November/December 2022 | CC-MAIN-2022-49 |

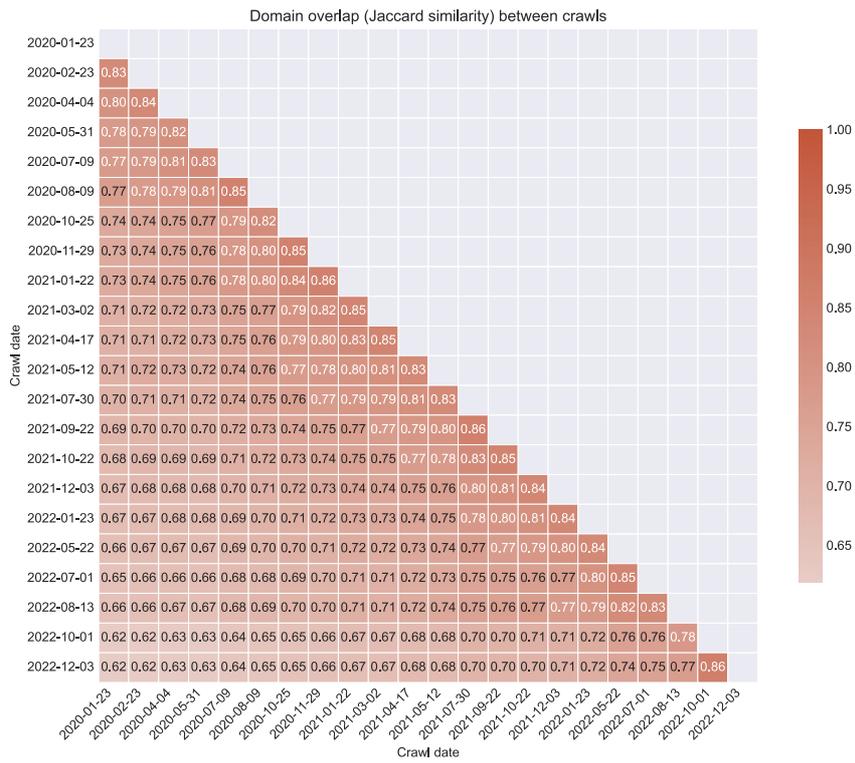

**Extended Data Figure 7**: Domain overlap between crawls (Jaccard similarity).

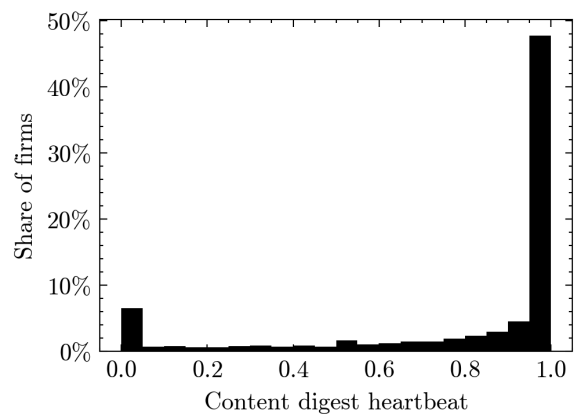

**Extended Data Figure 8**: Histogram of content digest heartbeat among firms. A value of 0 means that the firm's website content never changes between crawls, while a value of 1 means the content changes between every crawl.

**Llama Prompt**

```
"""You are given a text extract from a firm's website that is related to Covid-19. The text may be in a non-English
language. Assume that it is written from the firm's perspective unless otherwise specified. Your task is to analyze
the text and return  the information in the following format:

{
  affected: number, // Score the impact of Covid-19 on the firm as indicated by the text:
                    // 0: No indication of impact, only general pandemic information.
                    // 1: Slightly affected.
                    // 2: Moderately affected.
                    // 3: Significantly affected (e.g., closures or major operational changes).

  affectedness_category: string, // Categories indicating how the firm was affected:
                                 // Use one or more of {production, demand, supply}.
                                 // - Production: related to operations and employees.
                                 // - Supply: related to procurement and supply chains.
                                 // - Demand: related to customers.
                                 // Separate multiple categories with commas.

  tags: string // Tags describing specific ways the firm was affected:
               // Examples: closure of facilities, supply chain issues, home office implementation, customer hygiene
               measures.
               // Add new tags as appropriate. Separate multiple tags with commas.
}

Example paragraphs with expected output:

Input: "Beckhoff is reintroducing reinforced security measures due to the COVID19 infection. From Monday, October 26,
2020, the tried and tested two-shift system in production and an 80/20 home office rule for office workplaces will
apply again."
"Output: {"affected": 2, "affectedness_category": "production", "tags": "shift system, home office"}

Input: "Dear customers, due to the uncertainty about the development of the pandemic, we are closing the restaurant,
the shop and the reception until the end of March."
Output: {"affected": 3, "affectedness_category": "production, demand", "tags": "closure"}

Input: "The measures ordered by the Federal Council to contain the coronavirus pandemic and the associated current
situation are a challenge for everyone. We strive to maintain our operations and our services. We have no influence on
foreign suppliers if material is retained or blocked at the border, despite other statements in the media. We therefore
regret if some products are not available as a result."
Output: {"affected": 2, "affectedness_category": "supply", "tags": "supply chain issues, products unavailable"}

Input: "The health and safety of our employees and candidates is very important to us. We are closely monitoring
COVID-19 and have adjusted our recruiting procedures as needed. Peabody has adopted virtual recruiting tools, including
telephone and video interviews. This will allow us to meet new candidates and continue focus on bringing in top talent."
Output: {"affected": 1, "affectedness_category": "production", "tags": "recruiting procedures"}

Input: "Over the last five or six months we shared a series of wellbeing articles to support people during lockdown.
One focused on the benefits of physical activity, which we then backed up with our own intercompany activity challenge."
Output: {"affected": 0, "affectedness_category": "", "tags": ""}

Please output the extracted information in JSON format, following the provided schema. Do NOT add any clarifying
information. Output MUST follow the schema above. Do NOT add any additional output that does not appear in the schema.

Input: <Input paragraph>
Output: """
```

**Extended Data Figure 9**: Llama prompt.

| Tag Construction |
|---|
| **Supply chain issues**: supply chain, products unavailable, delivery, delay, logistics, product availability. |
| **Closure**: closed, closure, shutdown. |
| **Remote work**: home, remote, telecommut, digital, telework, virtual, online, smart working, Microsoft Teams, WFH, work-from-home, distance work, VPN, flexible work arrangements, flexible working. |
| **Hygiene measures**: quarantin, social distanc, hygiene, safety measures, PPE, hand sanitizer, health and safety, temperature, disinfect, mask, cleaning, employee safety, employee health, testing, vaccin, distancing, isolation, tracing, hand washing, sanitization, prevention, gloves, shift system, health measures, distance measure, sanitation, safety protocols, cleaning, personal protective equipment, video conferencing, screening, security measures, precautionary measures, infect, face coverings, preventive measures, face shields, customer safety, safety precautions, visitor restrictions, precautions, employee protection, access restrictions, curbside pickup, sanitizing, hand washing, access control, workplace safety, cleanliness, hand sanitiser, protective equipment, protective measures, appointment only, limited capacity, contact restrictions, test requirements, health measures, test requirement, PCR test, 3G, 2G, 3-G, 2-G. |
| **Travel restrictions**: travel. |
| **Financial impact**: revenue decrease, revenue decline, cost reduction, restructuring, liquidity, production halt, salary reduction, reduced demand, reduced sales, economic impact, layoffs, cash flow, production capacity, reduced hours, project delays, financial hardship, cost savings, financial difficulties, inventory management, job loss, revenue loss, sales decline, production suspension, cost-cutting, reduced workforce, reduced capacity, revenue impact, cost control, limited staff, bankruptcy, production slowdown, reduced revenue, reduced operations, event cancel, reduced staff, operational changes, Kurzarbeit, cancellations, events cancel, workforce reduction, shift system, event cancellation, payment deferral, cancelled events, event postponement, cancellation of events, event restrictions, cancelled event, insolvency communication, lockdown, COVID-19, customers, technology, support, recovery, flexibility, uncertainty, 3D printing, pandemic, security. |

**Extended Data Figure 10**: Tags proposed by the LLM that fall under the umbrella terms supply chain issues, closure, remote work, hygiene measures, travel restrictions and financial impact (matches are insensitive to case and word boundaries).

# Supplementary Information (SI) for
# "A Web-Based Indicator for Real-Time Monitoring of Economic Shocks"

Michael D. König[1,2,3], Jakob Rauch[1] and Martin Wörter[1]

[1]ETH Zurich, Swiss Economic Institute (KOF), Leonhardstrasse 21, Zurich, 8092, Switzerland.
[2]Tinbergen Institute and Dep. of Spatial Economics, VU Amsterdam, De Boelelaan 1105, Amsterdam, 1081 HV, The Netherlands.
[3] Centre for Economic Policy Research (CEPR), Great Sutton St. 33, London, EC1V 0DX, United Kingdom.

# Contents

# A Supplementary Tables and Figures: Sample construction and keywords

## A.1 Sample Coverage and Representativeness

Table A.1 shows the coverage of firms in our data sample for each country. For the US we cover 95% of firms with at least 5 employees, and typically at least half of such firms in EU countries. For countries outside of the OECD precise numbers on the total population of such companies is harder to obtain. We therefore also show in Table A.1 the number of firms analyzed per million inhabitants. We exclude countries with insufficient coverage in our data, either less than 1,000 analyzed firms per million inhabitants or less than 20 percent of firms covered. Note that the number of analyzed firms in each country in our data sample depends on the number of firms with employee, sector, and website data available in the Orbis database,[31] as well as the number of websites available for these firms in CommonCrawl.

The comprehensiveness and global coverage of WAI is illustrated in Figure A.1, which depicts the share of firms mentioning Covid-19 on their website across cities worldwide. The figure aggregates WAI scores at the city level, calculating the fraction of affected firms relative to the total number of firms in each location. Approximately 25% of all firms analyzed mention Covid-19 on their websites, with the largest cities in North America, Europe, East Asia, and South America showing up to 40% of firms referencing the pandemic. The figure highlights the extensive coverage of WAI, encompassing the major economies of the developed world.

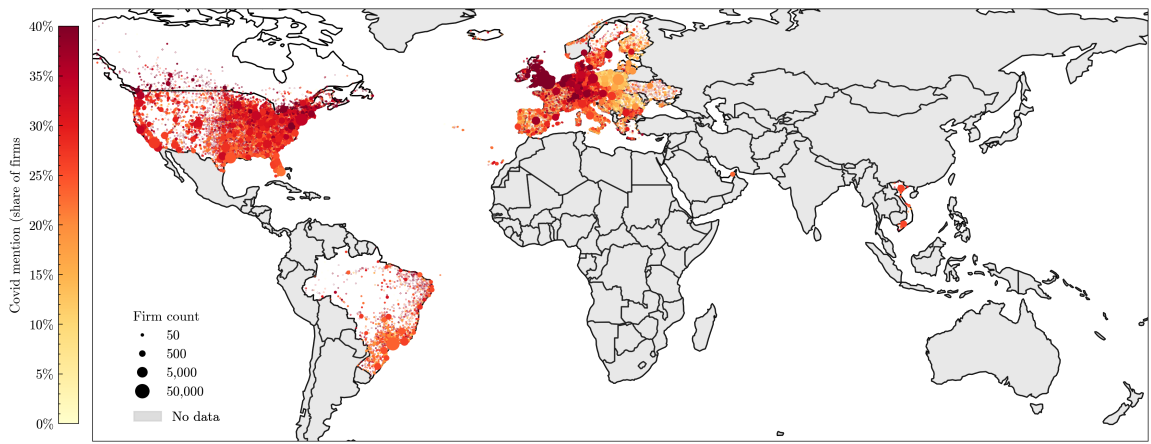

**Figure A.1**: The share of firms mentioning Covid-19 on their website at the city level, based on a sample of 5,429,830 companies with at least 5 employees.

Table A.1: Firm sample country coverage.

| Country | Firms analyzed | Firms analyzed per million inhabitants | Total firms (excluding <5 employees) | Share of firms analyzed (%) | Source of firm demographics data | Share of population using internet (%, OECD 2019) |
|---|---|---|---|---|---|---|
| United States of America | 2,193,914 | 6,736 | 2,312,130 | 95 | U.S. Census Bureau (2020) | 89 |
| United Kingdom | 381,326 | 5,776 | 557,370 | 68 | UK Statistics Authority (2020) | 93 |
| Germany | 376,714 | 4,555 | 610,070† | 62 | OECD (2020) | 88 |
| Russian Federation | 261,519 | 1,815 | | | Federal State Statistics Service (2015) | 83 |
| Brazil | 246,810 | 1,179 | 794,990† | 31 | OECD (2020) | 74 |
| Japan | 199,468 | 1,567 | 1,204,730* | 17 | Statistical Office of Japan (2014) | 93 |
| Italy | 194,751 | 3,216 | 372,347† | 52 | OECD (2020) | 68 |
| China | 186,861 | 135 | | | State Administration for Market Regulation (2017) | 64 |
| Poland | 185,042 | 4,873 | 215,855† | 86 | OECD (2020) | 80 |
| Spain | 131,047 | 2,814 | 285,410† | 46 | OECD (2020) | 91 |
| Netherlands | 98,707 | 5,827 | 108,148† | 91 | OECD (2020) | 93 |
| France | 93,788 | 1,397 | 329,489† | 28 | OECD (2020) | 83 |
| Canada | 73,368 | 1,999 | 298,350† | 25 | OECD (2020) | 92 |
| Sweden | 57,129 | 5,674 | 82,443† | 69 | OECD (2020) | 94 |
| Czech Republic | 49,810 | 4,703 | 69,147† | 72 | OECD (2020) | 81 |
| Austria | 49,165 | 5,581 | 75,819† | 65 | OECD (2020) | 88 |
| Belgium | 45,370 | 3,990 | 56,780† | 80 | OECD (2020) | 90 |
| Portugal | 39,320 | 3,820 | 94,966† | 41 | OECD (2020) | 75 |
| Norway | 38,118 | 7,344 | 49,016† | 78 | OECD (2020) | 98 |
| Republic of Korea | 38,021 | 745 | 853,449 | 4 | Statistics Korea (2019) | 96 |
| Switzerland | 37,222 | 4,494 | 97,783† | 38 | OECD (2020) | 93 |
| Denmark | 35,950 | 6,276 | 44,388† | 81 | OECD (2020) | 98 |
| Hungary | 34,563 | 3,534 | 62,473† | 55 | OECD (2020) | 80 |
| Finland | 30,610 | 5,570 | 39,779† | 77 | OECD (2020) | 90 |
| Romania | 27,151 | 1,386 | 104,468† | 26 | OECD (2020) | 74 |
| Vietnam | 27,023 | 289 | 138,478* | 20 | Statistical Yearbook of Vietnam 2016 (2015) | 69 |
| Mexico | 25,975 | 212 | 516,720† | 5 | OECD (2020) | 70 |
| Lithuania | 21,974 | 7,771 | 27,179† | 81 | OECD (2020) | 82 |
| Bulgaria | 20,864 | 2,949 | 48,423† | 43 | OECD (2020) | 68 |
| Slovakia | 19,528 | 3,596 | 30,634† | 64 | OECD (2020) | 83 |
| Ireland | 17,683 | 3,719 | 39,456† | 45 | OECD (2020) | 87 |
| Ukraine | 17,421 | 389 | 60,154* | 29 | State Statistics Service Ukraine (2017) | 70 |
| Greece | 15,815 | 1,470 | 79,868† | 20 | OECD (2020) | 76 |
| Croatia | 12,753 | 3,091 | 29,667† | 43 | OECD (2020) | 79 |
| Turkey | 12,672 | 157 | 347,210† | 4 | OECD (2020) | 74 |
| Serbia | 12,536 | 1,750 | 11,403* | 110 | OECD (2013) | 77 |
| Latvia | 9,686 | 4,991 | 18,023† | 54 | OECD (2020) | 86 |
| Slovenia | 9,577 | 4,634 | 16,556† | 58 | OECD (2020) | 83 |
| Estonia | 9,410 | 7,143 | 14,095† | 67 | OECD (2020) | 90 |
| Israel | 8,892 | 1,040 | 66,367† | 13 | OECD (2020) | 87 |
| Peru | 6,363 | 198 | 85,672* | 7 | Ministerio de la Produccion (2016) | 60 |
| Australia | 5,490 | 223 | 249,812 | 2 | Australian Bureau of Statistics. Counts of Australian Businesses (2017) | 94 |

**Notes:** Countries with less than 1,000 analyzed firms per million inhabitants or where less than 20 percent of all firms are analyzed (marked in red) are removed from further analysis. * indicates where the source data only gives total number of firms excluding firms with 0 to 10 employees (instead of 0 to 5). † indicates where OECD source data only gives total number of firms in the business economy (NACE codes 05 to 82 less 642).

Table A.1 continued.

| Country | Firms analyzed | Firms analyzed per million inhabitants | Total firms (excluding <5 employees) | Share of firms analyzed (%) | Source of firm demographics data | Share of population using internet (%, OECD 2019) |
|---|---|---|---|---|---|---|
| Colombia | 4,650 | 95 | 151,929† | 3 | OECD (2020) | 65 |
| United Arab Emirates | 4,285 | 622 | 20,233* | 21 | Agency Department of Economic Development (2008) | 99 |
| India | 4,019 | 3 | | | Ministry of Micro (2015) | 30 |
| New Zealand | 3,615 | 754 | 51,027† | 7 | OECD (2020) | 90 |
| Republic of Moldova | 3,556 | 1,002 | | | National Bureau of Statistics of the Republic of Moldova (2017) | 58 |
| Bosnia and Herzegovina | 3,516 | 1,003 | 8,660* | 41 | Agency for Statistics of Bosnia and Herzegovina (2017) | 70 |
| Belarus | 3,408 | 358 | | | National Statistical Committee of the Republic of Belarus (2017) | 83 |
| Iceland | 2,414 | 7,030 | 3,353† | 72 | OECD (2020) | 100 |
| Chile | 2,212 | 125 | 234,125* | 1 | SII Servicio de Impuestos Internos (2015) | 85 |
| Saudi Arabia | 2,092 | 64 | | | Small and Medium-Sized Establishments Survey 2017 (2017) | 96 |
| Ecuador | 1,989 | 120 | 80,109* | 2 | INE. Instituto Nacional de Estadisticas y censos. Directorio de Empresas (2016) | 59 |
| Egypt | 1,797 | 18 | 432,737 | 0 | CAPMAS Establishment Census 2017 (2017) | 57 |
| Sri Lanka | 1,789 | 86 | 83,945* | 2 | Department of Census and Statistics Sri Lanka (2014) | 32 |
| Cayman Islands | 1,665 | | | | | |
| Georgia | 1,495 | 402 | | | National Statistics Office of Georgia (2017) | 69 |
| North Macedonia | 1,196 | 575 | 3,295* | 36 | Eurostat (2016) | 81 |
| Uzbekistan | 1,133 | 35 | | | The State Committee of the Republic of Uzbekistan on Statistics (2018) | 70 |
| Luxembourg | 1,115 | 1,860 | 8,877† | 13 | OECD (2020) | 97 |
| Montenegro | 1,063 | 1,709 | | | Statistical Office of Montenegro (2017) | 73 |
| Pakistan | 980 | 6 | | | PFBS. Economic Census 2005 (2005) | 17 |
| Malaysia | 961 | 31 | 226,954 | 0 | Economic Census 2015 (2015) | 84 |
| Kazakhstan | 862 | 48 | | | Committee on Statistics (2016) | 82 |
| Uruguay | 843 | 244 | 29,095 | 3 | INE (2016) | 83 |
| Indonesia | 763 | 3 | 821,177 | 0 | Ministry of Cooperatives and SMEs Indonesia Government (2017) | 48 |
| Bermuda | 752 | 11,491 | 1,036 | 73 | Department of Statistics (2017) | |
| Kuwait | 730 | 331 | 24,750* | 3 | WAMDA (2004) | 100 |
| Liechtenstein | 678 | 17,879 | 556* | 122 | Landesverwaltung Fürstentum Liechtenstein (2017) | 93 |

**Notes:** Countries with less than 1,000 analyzed firms per million inhabitants or where less than 20 percent of all firms are analyzed (marked in red) are removed from further analysis. * indicates where the source data only gives total number of firms excluding firms with 0 to 10 employees (instead of 0 to 5). † indicates where OECD source data only gives total number of firms in the business economy (NACE codes 05 to 82 less 642).

## A.2 Covid-19 Synonyms

We extracted paragraphs from the company websites containing at least one of a list of Covid-related keywords translated into 65 languages. This was done using the following prompt for the large language model GPT-3.5-turbo:

> **Translation**
> Translate the following keywords into different languages: 'corona', 'covid', 'covid-19', 'sars-cov-2', 'coronavirus', 'pandemic'. Please only output the translated keywords, each on a new line.

The resulting keywords can be found in Table A.2.

| Language | Keywords |
| --- | --- |
| Albanian | korona, covid, covid-19, sars-cov-2, koronaviru... |
| Arabic | كورونا, كوفيد, كوفيد-19, فيروس كوف-2, سارس-كوف... فيروس |
| Armenian | քորոնա, covid, covid-19, sars-cov-2, քորոնավիրը... |
| Azerbaijani | korona, kovid, kovid-19, sars-cov-2, koronaviru... |
| Basque | korona, covid, covid-19, sars-cov-2, koronabiru... |
| Belarusian | каранавірус, кавід, кавід-19, sars-cov-2, каран... |
| Bengali | করোনা, কোভিড, কোভিড-১৯, সার্স-কোভ-২, করোনাভাইরাস... |
| Bosnian | korona, covid, covid-19, sars-cov-2, koronaviru... |
| Bulgarian | корона, ковид, ковид-19, сарс-ков-2, коронавиру... |
| Burmese | ကိုရိုနာ, ကိုဗစ်, ကိုဗစ်-၁၉, စက်မှုရောဂါ-၂,... |
| Catalan | corona, covid, covid-19, sars-cov-2, coronaviru... |
| Chinese | 冠状病毒, covid, covid-19, sars-cov-2, 新型冠状病毒, 大流行病... |
| Croatian | korona, covid, covid-19, sars-cov-2, koronaviru... |
| Czech | korona, covid, covid-19, sars-cov-2, koronaviru... |
| Danish | corona, covid, covid-19, sars-cov-2, coronaviru... |
| Dutch | corona, covid, covid-19, sars-cov-2, coronaviru... |
| English | corona, covid, covid-19, sars-cov-2, coronaviru... |
| Estonian | koroonaviirus, covid, covid-19, sars-cov-2, kor... |
| Finnish | korona, covid, covid-19, sars-cov-2, koronaviru... |
| French | corona, covid, covid-19, sars-cov-2, coronaviru... |
| Galician | coroa, covid, covid-19, sars-cov-2, coronavirus... |
| Georgian | კორონა, კოვიდი, კოვიდ-19, sars-cov-2, კორონავირ, ... |
| German | corona, covid, covid-19, sars-cov-2, coronaviru... |
| Gujarati | કોરોના, કોવિડ, કોવિડ-19, સાર્સ-કોવ-2, coronavírus... |
| Hebrew | הקורונה נגיף סארס-קוב-2, קוביד-19, קוביד, קורונה |
| Hindi | कोरोना, कोविड, कोविड-१९, सार्स-कोव-२, कोरोनावायरस |
| Hungarian | corona, covid, covid-19, sars-cov-2, coronavíru... |
| Icelandic | kórónuveira, covid, covid-19, sars-cov-2, koron... |
| Indonesian | corona, covid, covid-19, sars-cov-2, coronaviru... |
| Italian | corona, covid, covid-19, sars-cov-2, coronaviru... |
| Japanese | コロナ, コビッド, コビッド-19, sars-cov-2, コロナウイルス, パンデミック |

**Table A.2**: Covid-19 synonyms.

| Language | Keywords |
| --- | --- |
| Kannada | ಕೊರೊನಾ, ಕೋವಿಡ್, ಕೋವಿಡ್-19, ಸಾರ್ಸ್-ಕೋವ್-2, ಕೊರೊನಾ ವೈರಸ್ |
| Kazakh | корона, ковид, ковид-19, сарс-ков-2, коронавирустың |
| Korean | 코로나, 코비드, 코비드-19, 사스-코프-2, 코로나바이러스, 팬데믹 |
| Latin | corona, covid, covid-19, sars-cov-2, coronaviru... |
| Latvian | korona, covid, covid-19, sars-cov-2, koronavīru... |
| Lithuanian | corona, covid, covid-19, sars-cov-2, coronaviru... |
| Macedonian | корона, ковид, ковид-19, сарс-ков-2, коронавирус |
| Malay | corona, covid, covid-19, sars-cov-2, coronaviru... |
| Malayalam | കൊറോണ, covid-19, സാറ്‍സ്-കോവ-2, കൊറോണവ... |
| Marathi | कोरोना, कोविड, कोविड-19, सार्स-कोव-2, कोरोनाव्हायरस |
| Modern Greek | κορώνα, κοβίντα, κοβίντ-19, σαρς-κοβ-2, κορονοϊ... |
| Mongolian | корона, ковид, ковид-19, сарс-ков-2, коронавируст |
| Nepali | कोरोना, कोभिड, कोभिड-१९, एसएआरएस-कोवी-२, कोरोना |
| Norwegian | korona, covid, covid-19, sars-cov-2, koronaviru... |
| Polish | koronawirus, covid, covid-19, sars-cov-2, koron... |
| Portuguese | corona, covid, covid-19, sars-cov-2, coronavíru... |
| Romanian | corona, covid, covid-19, sars-cov-2, coronaviru... |
| Russian | корона, ковид, ковид-19, сарс-ков-2, коронавиру... |
| Serbian | korona, kovid, kovid-19, sars-kov-2, koronaviru... |
| Slovak | korona, covid, covid-19, sars-cov-2, koronavíru... |
| Slovenian | korona, covid, covid-19, sars-cov-2, koronaviru... |
| Spanish | corona, covid, covid-19, sars-cov-2, coronaviru... |
| Swedish | corona, covid, covid-19, sars-cov-2, coronaviru... |
| Tamil | கொரோனா, கோவிட், கோவிட்-19, சார்ஸ்-கோவ்-2, கொரோன... |
| Telugu | కరోనా, కోవిడ్, కోవిడ్-19, సార్స్-కోవ్-2, కరోనవై... |
| Thai | โคโรนา, โควิด, โควิด-19, ซาร์ส-โควี-2, โคโรไนป์... |
| Turkish | corona, covid, covid-19, sars-cov-2, coronavirü... |
| Ukrainian | корона, ковід, ковід-19, сарс-ков-2, коронавірус |
| Urdu | کورونااوائرس, کوو-2, سارس کووڈ-19, کووڈ, کرونا, ... |
| Uzbek | korona, kovid, kovid-19, sars-cov-2, koronaviru... |
| Vietnamese | corona, covid, covid-19, sars-cov-2, coronaviru... |
| Welsh | corona, covid, covid-19, sars-cov-2, coronafeir... |

**Table A.2** continued.

# B Supplementary Discussion: WAI geographic and temporal variation

In the following sections we provide a complete overview of the US states and countries considered in our analysis complementing Figures 2 and 3, respectively, in the main text. In particular, in Section B.1 we show the Covid-19 affectedness based on WAI and OxCGRT among the fifty largest states in the US, while in Section B.2 we provide an international comparison.

## B.1 Variation Across US States

Figure B.1 shows the share of firms mentioning Covid-19 and the severity of affectedness by US states. For each firm, we take the maximum degree of affectedness over the period of analysis. The most affected states are the District of Columbia, Vermont and Maine, while the least affected states are Florida, Utah and Georgia. Similar to Figure 2 in the main text, we can analyze variations of affectedness also over time and space. Figures B.2, B.3 and B.4 show the Covid-19 affectedness based on WAI and OxCGRT among US states (in terms of the number of companies analyzed) together with the number of analyzed firms, and the Pearson correlation coefficient ($r$) between WAI and OxCGRT (in the first differences of the two time series). For almost all states the correlation coefficient, $r$, between WAI and OxCGRT is above 0.8, indicating that at the aggregate state level both measures are highly correlated. The spatial maps show that the population centers in each state exhibit the largest fractions of affected firms.

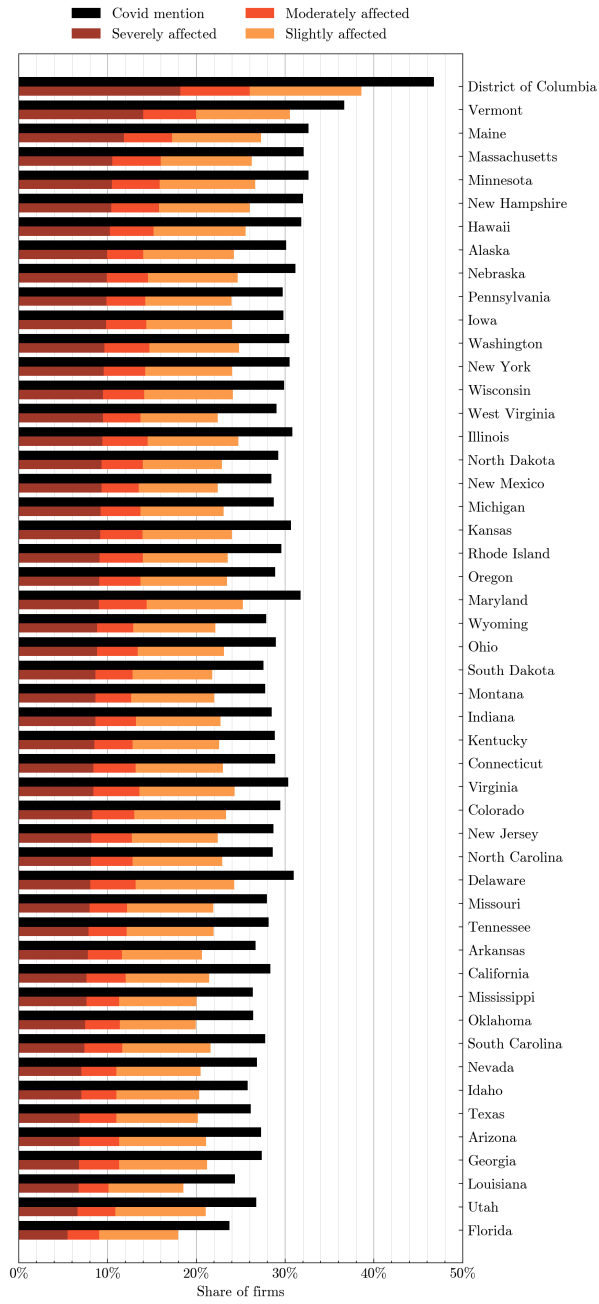

**Figure B.1**: Share of firms mentioning Covid on their website, as well as the intensity of affectedness based on company websites for US states.

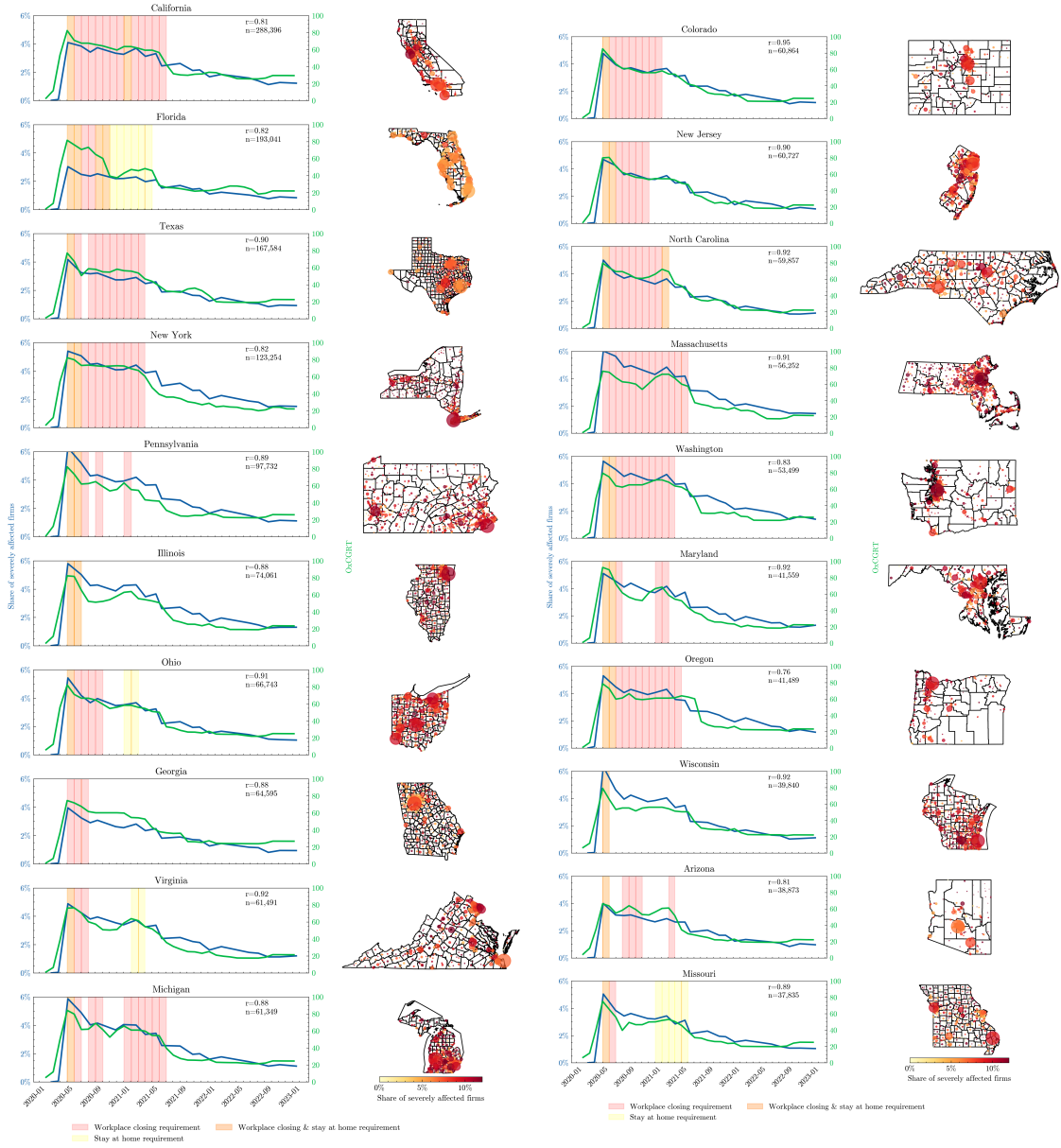

**Figure B.2**: Left panels show Covid-19 affectedness based on WAI and OxCGRT among the twenty largest states in the US (in terms of the number of companies analyzed) together with the number of analyzed firms ($n$), and the Pearson correlation coefficient ($r$) between WAI and OxCGRT (first differences). Time periods in which workplace closing or stay at home requirements were implemented are indicated with vertical bars. Right, spatial maps indicating WAI at the city level.

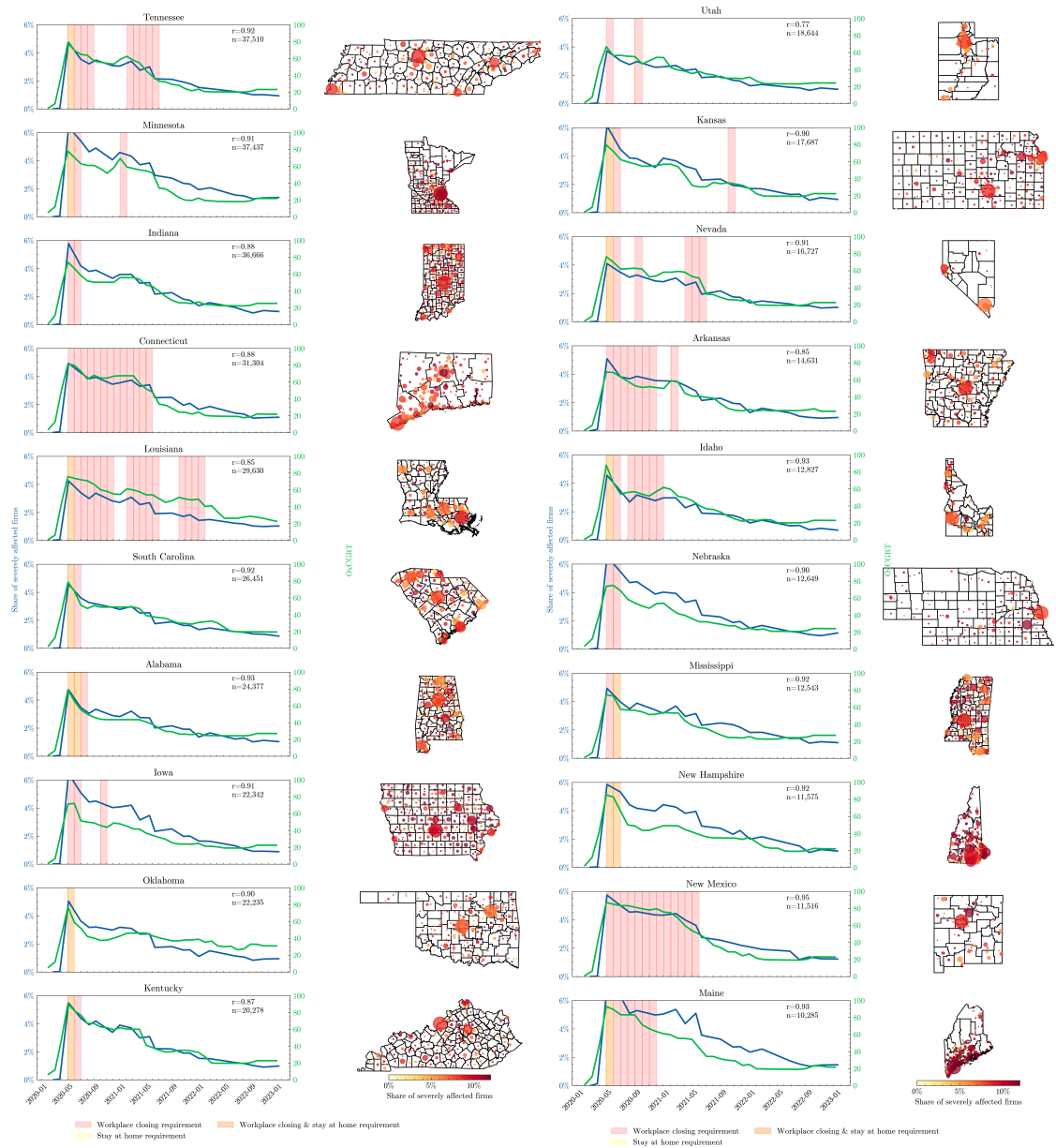

**Figure B.3**: Left panels show Covid-19 affectedness based on WAI and OxCGRT among the twenty one to forty largest states in the US (in terms of the number of companies analyzed) together with the number of analyzed firms (*n*), and the Pearson correlation coefficient (*r*) between WAI and OxCGRT (first differences). Time periods in which workplace closing or stay at home requirements were implemented are indicated with vertical bars. Right, spatial maps indicating WAI at the city level.

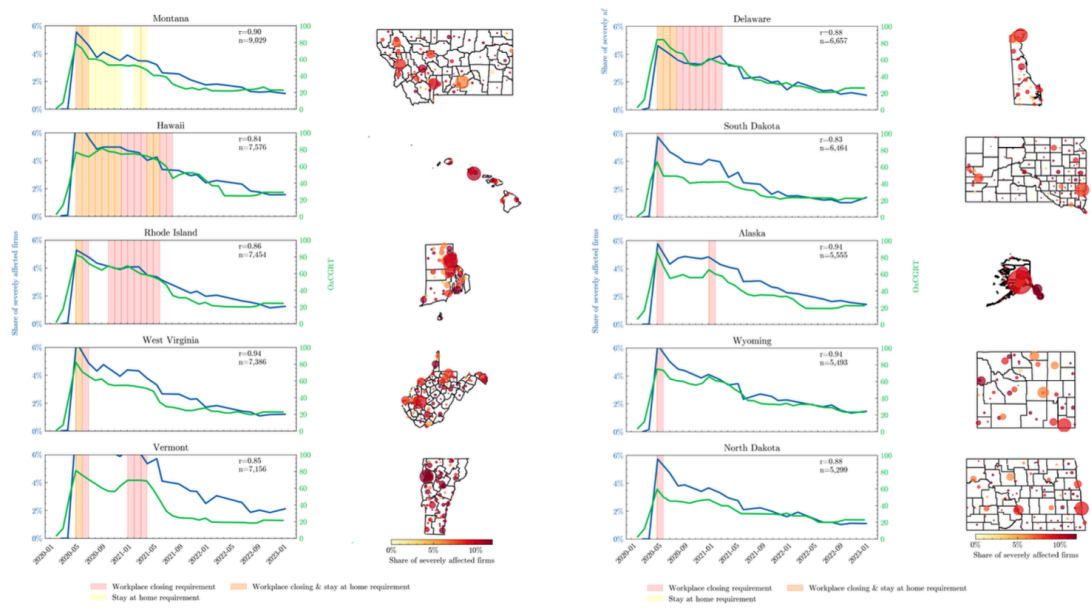

**Figure B.4**: Left panels show Covid-19 affectedness based on WAI and OxCGRT among the forty one to fiftieths largest states in the US (in terms of the number of companies analyzed) together with the number of analyzed firms ($n$), and the Pearson correlation coefficient ($r$) between WAI and OxCGRT (first differences). Time periods in which workplace closing or stay at home requirements were implemented are indicated with vertical bars. Right, spatial maps indicating WAI at the city level.

## B.2 International Comparison

In this section we provide a comparison of WAI across countries. For this purpose we only consider countries where the share of the population using the internet is above the EU average (of 82.76%) in the year 2019 using data from the International Telecommunication Union.[33] The motivation for this threshold is that websites are expected to be used by firms as an effective communication tool (which is a prerequisite for WAI) only if the internet usage in the population is sufficiently high.[19]

Figure B.5 shows the share of firms mentioning Covid-19 together with each affectedness category across countries. For each firm analyzed we take the maximum degree of affectedness over the period of analysis. We see strong differences across countries, with strongly affected countries including the UK, Canada, Belgium, and less affected countries including Nordic and Eastern European countries such as Latvia, Estonia or Finland. Moreover, similar to Figure 3 in the main text, we can analyze variations of affectedness also over time and space. Figure B.6 shows Covid-19 affectedness based on WAI and OxCGRT among the largest countries (in terms of the number of companies analyzed) together with the number of analyzed firms, and the Pearson correlation coefficient between WAI and OxCGRT (of the first differences of the two time series). Figure B.6 shows that for almost all countries analyzed the correlation coefficient between WAI and OxCGRT is above 0.56, indicating that at the aggregate country level both measures are highly correlated. The spatial maps show that the population centers in each country exhibit the largest fractions of affected firms.

We note that the correlation is slightly lower than what we have observed for the US states in the previous Section B.1. The cross-country comparison seems to suggests that WAI performs best in countries with widespread internet usage (cf. Table A.1) and a predominantly English-speaking population, such as the US, Canada, the UK, and Ireland. A possible explanation could be that the LLMs examined in this study are primarily trained on English-language corpora, which may give them an advantage in these countries. However, with increasing digitization across countries and improvements in multi-lingual LLMs we expect these differences to become less prominent over time, and thus making WAI an even more valuable tool in the future.

---

[19]The complete set of all countries can be accessed via the public dashboard available at: https://covid-explorer.kof.ethz.ch/.

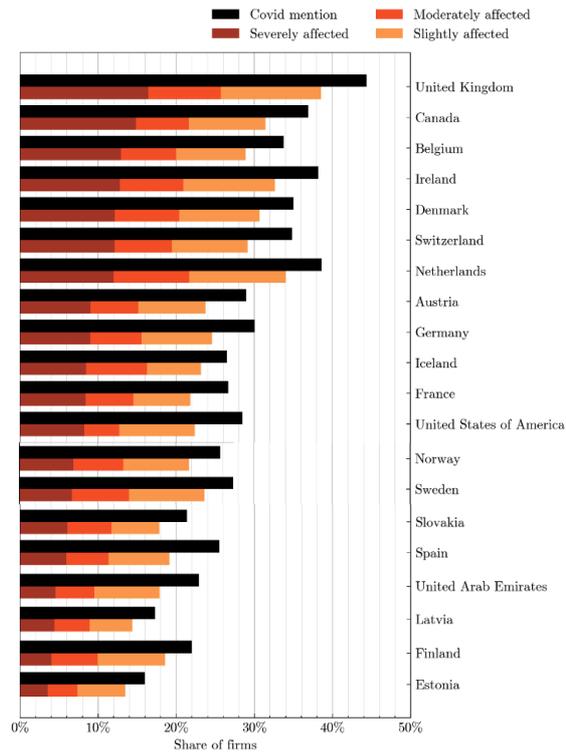

**Figure B.5**: Share of firms mentioning Covid-19 on their website, as well as the intensity of affectedness based on website text for the top 50 countries by number of analyzed firms. We only consider countries where the share of the population using the internet is above the EU average in the year 2019.

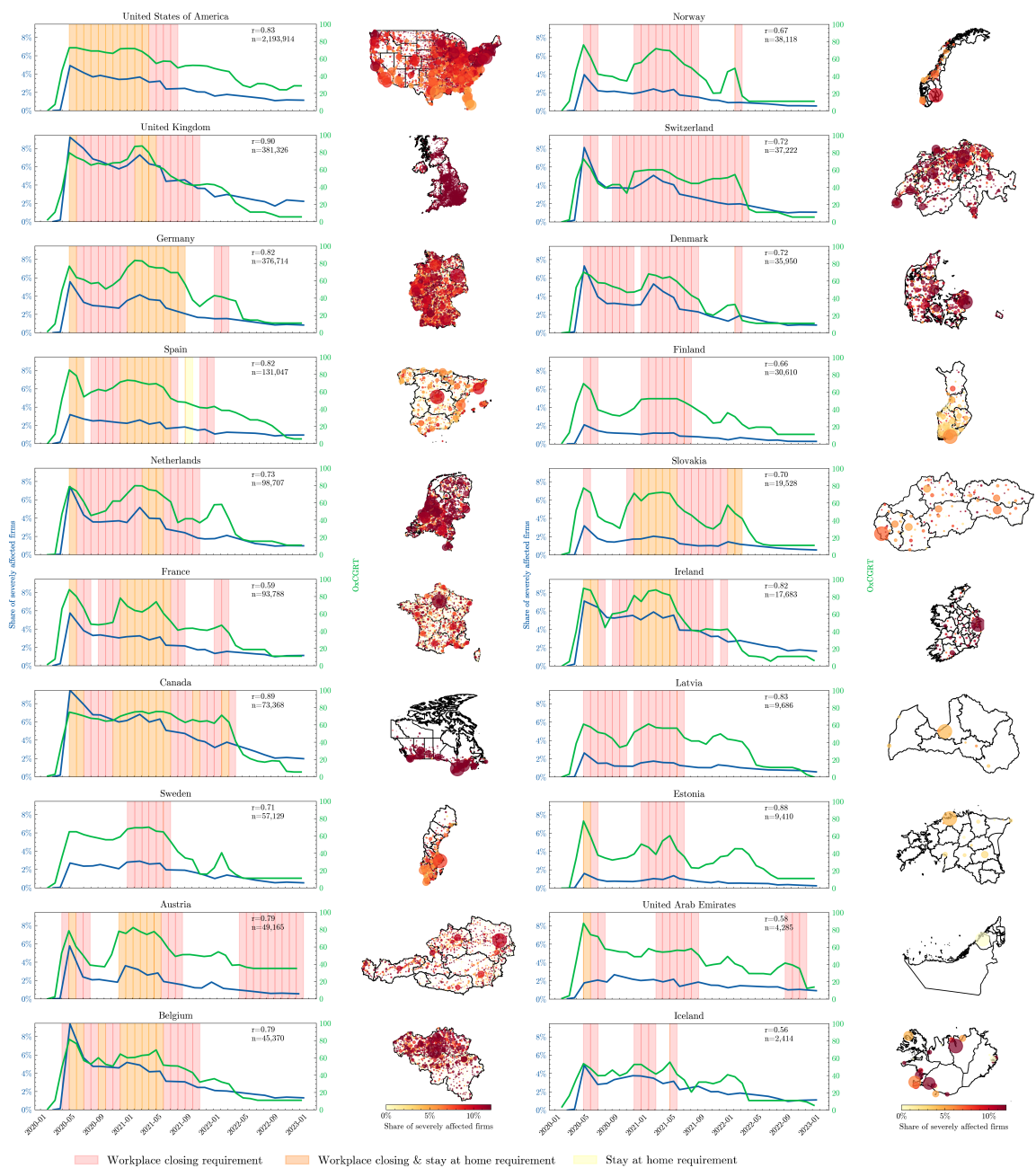

**Figure B.6**: Left panels show Covid-19 affectedness based on WAI and OxCGRT among the largest countries (in terms of the number of companies analyzed) together with the number of analyzed firms ($n$), and the Pearson correlation coefficient ($r$) between WAI and OxCGRT. Time periods in which workplace closing or stay at home requirements were implemented are indicated with vertical bars. Right, spatial maps indicating WAI at the city level. We only consider countries where the share of the population using the internet is above the EU average in the year 2019.

# C Supplementary Notes: Estimation and robustness

In the following sections we present the details of the estimation results shown in Figure 4 in the main text and provide additional robustness checks for different LLMs, time horizons, sectors and specifications. Specifically, Section C.1 shows the estimation results across the different specifications in Eqs. (1), (2) and (3). Next, Section C.2 presents several robustness checks, where in Section C.2.1 we use an alternative LLM (ChatGPT) to compute WAIs, in Section C.2.2 we estimate the model for different time periods, in Section C.2.3 we estimate the model for the manufacturing and services sectors separately, and in Section C.2.4 we estimate a model that includes a one-period lag of the dependent variable to account for potential serial correlation.

## C.1 Estimation Results

Table C.1 presents the complete set of estimation results for Eqs. (1) to (3), using quarterly sales growth (in percentages) in columns (1) to (3), and stock returns (in percentages) in columns (4) to (6) as the dependent variables, covering the period from 2017 to 2022. The independent variables include the WAI Covid-19 affectedness categories based on the LLM. A graphical illustration of annualized coefficients from these estimates is provided in Figure 4 in the main text.

In particular, in columns (1) and (4) in Table C.1 correspond to the estimation Equation (1) with the Covid-19 mentions and the WAI Covid-19 affectedness categories – mild, moderate, and severe – based on the Llama 3.1 LLM output as explanatory variables. Columns (2) and (5) in Table C.1 correspond to the estimation Equation (2) with additional controls from OxCGRT. The controls included are workplace closing recommended or required for all-but-essential sectors, stay at home recommended or required with minimal exceptions, log fiscal measures, and the number of deaths from Covid-19 per million. As expected, these policy measures have negative signs for sales growth, except for fiscal measures which are assisting companies directly or indirectly. However, we find that the estimated coefficient changes sign when analyzing stock returns. This could be explained by the fact that stock prices reflect expectations for future earnings, if investors believe that taking such measures has long-term benefits that outweigh short-term damages. Columns (3) and (6) in Table C.1 correspond to the estimation Equation (3) where instead of the OxCGRT policy controls we include country-industry-quarter fixed effects. In all columns we include lagged log total assets, firm fixed effects and quarter fixed effects.

Form the estimation results in Table C.1 we find that WAIs have significant predictive power for sales growth rates and stock returns under all specifications, confirming that these indicators provide timely and relevant firm-level information on Covid-19 affectedness.

Table C.1: Regression results corresponding to Eqs. (1) to (3) for log quarterly sales growth and log quarterly stock returns as dependent variables on Llama 3.1 affectedness variables (mild, moderate and severe) as predictors (cf. Figure 4).

|  | Log sales growth (%) | | | Log stock returns (%) | | |
| --- | --- | --- | --- | --- | --- | --- |
|  | (1) | (2) | (3) | (4) | (5) | (6) |
| Covid mention | 0.390<br>(0.430) | 0.551<br>(0.432) | 0.717<br>(0.460) | -0.481**<br>(0.154) | -0.629***<br>(0.155) | -0.420**<br>(0.160) |
| Affected (Llama) | | | | | | |
|   Mild | -0.848<br>(0.693) | -0.777<br>(0.703) | -0.197<br>(0.888) | -0.628**<br>(0.223) | -0.743***<br>(0.224) | -0.753**<br>(0.239) |
|   Moderate | -3.180***<br>(0.578) | -3.080***<br>(0.599) | -2.022*<br>(0.790) | -0.982***<br>(0.223) | -1.120***<br>(0.226) | -0.843***<br>(0.250) |
|   Severe | -3.892***<br>(0.798) | -3.818***<br>(0.813) | -3.059**<br>(1.022) | -1.138***<br>(0.292) | -1.330***<br>(0.294) | -1.121***<br>(0.326) |
| Lagged log total assets | -5.843***<br>(0.663) | -5.759***<br>(0.664) | -5.713***<br>(0.755) | -4.386***<br>(0.445) | -4.365***<br>(0.448) | -4.393***<br>(0.194) |
| Workplace closing | | | | | | |
|   Recommended | | -4.552***<br>(0.513) | | | 4.109***<br>(0.176) | |
|   Required | | -6.339***<br>(0.610) | | | 6.268***<br>(0.202) | |
| Stay at home requirements | | | | | | |
|   Recommended | | -2.583***<br>(0.488) | | | 2.481***<br>(0.161) | |
|   Required | | -1.380*<br>(0.553) | | | 0.229<br>(0.194) | |
| Log fiscal measures | | 0.148***<br>(0.0315) | | | -0.0185*<br>(0.00934) | |
| Covid deaths per month | | 0.0291<br>(0.0990) | | | -0.0566<br>(0.0344) | |
| Firm FE | Yes | Yes | Yes | Yes | Yes | Yes |
| Quarter FE | Yes | Yes | Yes | Yes | Yes | Yes |
| Country-Industry-Quarter FE | No | No | Yes | No | No | Yes |
| No. Firms | 32610 | 32480 | 30602 | 33030 | 32880 | 31281 |
| Observations | 613702 | 611435 | 573847 | 627840 | 625362 | 591592 |
| $R^2$ | 0.0411 | 0.0414 | 0.0994 | 0.118 | 0.121 | 0.300 |

Standard errors in parentheses.
Standard errors clustered at the firm level.
Significance levels indicated by: * $p < 0.05$, ** $p < 0.01$, *** $p < 0.001$.

## C.2 Robustness

In the following we provide additional estimation results illustrating the robustness of our findings. Section C.2.1 assess the robustness of our estimation results when using ChatGPT as LLM. Section C.2.2 considers alternative time periods, Section C.2.3 analyzes different sectors, and Section C.2.4 estimates the model including a lagged dependent variable as an additional covariate.

### C.2.1 ChatGPT LLM

To assess the robustness of our estimation results, we use an alternative LLM to compute WAIs. Table C.2 presents the results, with log quarterly sales growth (in percentages) in columns (1) to (3), and log quarterly stock returns (in percentages) in columns (4) to (6) as the dependent variables, spanning the period from 2017 to 2022. The WAI Covid-19 affectedness categories—mild, moderate, and severe—are used as predictors, computed using OpenAI's ChatGPT LLM. The other covariates are the same as in in Table C.1. When compared to the results in Table C.1, the estimates show a high degree of similarity.

Table C.2: Regression results corresponding to Eqs. (1) to (3) for log quarterly sales growth and log quarterly stock returns as dependent variables on ChatGPT affectedness variables (mild, moderate and severe) as predictors.

|  | Log sales growth (%) | | | Log stock returns (%) | | |
| --- | --- | --- | --- | --- | --- | --- |
|  | (1) | (2) | (3) | (4) | (5) | (6) |
| Covid mention | 0.570 (0.438) | 0.721 (0.440) | 0.817 (0.466) | -0.477** (0.157) | -0.615*** (0.157) | -0.421** (0.161) |
| Affected (ChatGPT) | | | | | | |
|   Mild | -1.712* (0.720) | -1.624* (0.730) | -1.125 (0.906) | -0.333 (0.233) | -0.449 (0.235) | -0.357 (0.250) |
|   Moderate | -2.520*** (0.571) | -2.415*** (0.590) | -1.321 (0.775) | -0.795*** (0.213) | -0.946*** (0.216) | -0.687** (0.238) |
|   Severe | -4.032*** (0.681) | -3.911*** (0.695) | -3.061*** (0.869) | -0.863** (0.271) | -1.024*** (0.273) | -0.735* (0.300) |
| Lagged log total assets | -5.839*** (0.664) | -5.756*** (0.665) | -5.712*** (0.756) | -4.389*** (0.445) | -4.368*** (0.448) | -4.397*** (0.194) |
| Workplace closing | | | | | | |
|   Recommended | | -4.551*** (0.513) | | | 4.102*** (0.176) | |
|   Required | | -6.361*** (0.610) | | | 6.255*** (0.202) | |
| Stay at home requirements | | | | | | |
|   Recommended | | -2.559*** (0.487) | | | 2.488*** (0.161) | |
|   Required | | -1.367* (0.553) | | | 0.238 (0.194) | |
| Log fiscal measures | | 0.147*** (0.0315) | | | -0.0187* (0.00934) | |
| Covid deaths per month | | 0.0305 (0.0989) | | | -0.0619 (0.0344) | |
| Firm FE | Yes | Yes | Yes | Yes | Yes | Yes |
| Quarter FE | Yes | Yes | Yes | Yes | Yes | Yes |
| Country-Industry-Quarter FE | No | No | Yes | No | No | Yes |
| No. Firms | 32610 | 32480 | 30602 | 33030 | 32880 | 31281 |
| Observations | 613702 | 611435 | 573847 | 627840 | 625362 | 591592 |
| $R^2$ | 0.0411 | 0.0414 | 0.0994 | 0.118 | 0.121 | 0.300 |

Standard errors in parentheses.
Standard errors clustered at the firm level.
Significance levels indicated by: * $p < 0.05$, ** $p < 0.01$, *** $p < 0.001$.

### C.2.2 Different Time Periods

Table C.3 provides estimation results for quarterly sales growth and stock returns for the periods 2018-2021 and Table C.4 for the periods 2019-2020, respectively. When estimating the model for different time periods, we find that the coefficients for WAI tend to be higher and less precisely measured when we consider the shorter sample periods from 2018 to 2021 or from 2019 to 2020, both, for quarterly sales growth and for stock returns. This is mainly driven by the smaller number of observations. This means that while the precision gets smaller, the estimated effect increases.

**Table C.3**: Regression results corresponding to Eqs. (1) to (3) for log quarterly sales growth and log quarterly stock returns as dependent variables over the periods 2018-2021.

|  | Log sales growth | | | Log stock returns | | |
|---|---|---|---|---|---|---|
|  | (1) | (2) | (3) | (4) | (5) | (6) |
| Covid mention | -0.209 | 0.176 | 0.194 | 0.0578 | -0.155 | -0.457* |
|  | (0.549) | (0.553) | (0.586) | (0.207) | (0.208) | (0.211) |
| Affected |  |  |  |  |  |  |
|   Mild | -2.032* | -1.573 | -0.320 | -0.502 | -0.733* | -0.877** |
|  | (0.865) | (0.877) | (1.116) | (0.283) | (0.286) | (0.300) |
|   Moderate | -3.666*** | -3.105*** | -1.590 | -0.651* | -0.916*** | -1.022*** |
|  | (0.686) | (0.715) | (0.943) | (0.272) | (0.276) | (0.306) |
|   Severe | -4.294*** | -3.794*** | -2.180 | -0.616 | -0.942** | -1.226** |
|  | (0.971) | (0.991) | (1.235) | (0.362) | (0.364) | (0.400) |
| Lagged log total assets | -6.434*** | -6.532*** | -6.222*** | -4.796*** | -4.715*** | -4.966*** |
|  | (1.006) | (1.010) | (1.106) | (0.728) | (0.733) | (0.317) |
| Workplace closing |  |  |  |  |  |  |
|   Recommended |  | -11.97*** |  |  | 4.569*** |  |
|  |  | (0.851) |  |  | (0.312) |  |
|   Required |  | -12.75*** |  |  | 6.930*** |  |
|  |  | (0.900) |  |  | (0.315) |  |
| Stay at home requirements |  |  |  |  |  |  |
|   Recommended |  | -1.199* |  |  | 3.105*** |  |
|  |  | (0.591) |  |  | (0.217) |  |
|   Required |  | 0.0610 |  |  | 1.319*** |  |
|  |  | (0.680) |  |  | (0.252) |  |
| Log fiscal measures |  | 0.172*** |  |  | -0.0224* |  |
|  |  | (0.0329) |  |  | (0.00957) |  |
| Covid deaths per month |  | 0.0826 |  |  | -0.130** |  |
|  |  | (0.113) |  |  | (0.0399) |  |
| Firm FE | Yes | Yes | Yes | Yes | Yes | Yes |
| Quarter FE | Yes | Yes | Yes | Yes | Yes | Yes |
| Country-Industry-Quarter FE | No | No | Yes | No | No | Yes |
| No. Firms | 31396 | 31268 | 29484 | 31537 | 31395 | 29866 |
| Observations | 417334 | 415743 | 390651 | 427805 | 426058 | 403480 |
| $R^2$ | 0.0538 | 0.0545 | 0.113 | 0.135 | 0.138 | 0.320 |

Standard errors in parentheses.
Standard errors clustered at the firm level.
Significance levels indicated by: * $p < 0.05$, ** $p < 0.01$, *** $p < 0.001$.

**Table C.4**: Regression results corresponding to Eqs. (1) to (3) for log quarterly sales growth and log quarterly stock returns as dependent variables over the periods 2019-2020.

|  | Log sales growth (%) | | | Log stock returns (%) | | |
| --- | --- | --- | --- | --- | --- | --- |
|  | (1) | (2) | (3) | (4) | (5) | (6) |
| Covid mention | -1.238 | 0.270 | -0.277 | 1.099** | 0.202 | 0.0287 |
|  | (0.944) | (0.950) | (1.003) | (0.339) | (0.341) | (0.346) |
| Affected |  |  |  |  |  |  |
|   Mild | -4.229** | -1.967 | -0.237 | 0.751 | -0.0371 | -0.298 |
|  | (1.351) | (1.395) | (1.772) | (0.491) | (0.501) | (0.527) |
|   Moderate | -4.672*** | -1.778 | 1.371 | 0.736 | -0.335 | -0.438 |
|  | (1.198) | (1.263) | (1.684) | (0.448) | (0.464) | (0.506) |
|   Severe | -7.181*** | -4.418** | -2.308 | -0.341 | -1.138 | -2.126** |
|  | (1.562) | (1.628) | (1.960) | (0.662) | (0.676) | (0.769) |
| Lagged log total assets | -12.20*** | -12.43*** | -12.41*** | -8.155*** | -7.756*** | -5.672*** |
|  | (2.306) | (2.319) | (2.752) | (0.744) | (0.731) | (0.719) |
| Workplace closing |  |  |  |  |  |  |
|   Recommended |  | -16.63*** |  |  | 6.706*** |  |
|  |  | (1.383) |  |  | (0.506) |  |
|   Required |  | -18.87*** |  |  | 10.90*** |  |
|  |  | (1.489) |  |  | (0.530) |  |
| Stay at home requirements |  |  |  |  |  |  |
|   Recommended |  | -1.717 |  |  | 5.106*** |  |
|  |  | (1.211) |  |  | (0.428) |  |
|   Required |  | 1.502 |  |  | -2.334*** |  |
|  |  | (1.309) |  |  | (0.487) |  |
| Log fiscal measures |  | 0.102* |  |  | -0.189*** |  |
|  |  | (0.0450) |  |  | (0.0146) |  |
| Covid deaths per month |  | -0.654** |  |  | -0.111 |  |
|  |  | (0.199) |  |  | (0.0701) |  |
| Firm FE | Yes | Yes | Yes | Yes | Yes | Yes |
| Quarter FE | Yes | Yes | Yes | Yes | Yes | Yes |
| Country-Industry-Quarter FE | No | No | Yes | No | No | Yes |
| No. Firms | 29288 | 29167 | 27523 | 29689 | 29553 | 28076 |
| Observations | 208903 | 208094 | 195617 | 214669 | 213784 | 202509 |
| $R^2$ | 0.102 | 0.104 | 0.166 | 0.199 | 0.207 | 0.378 |

Standard errors in parentheses.
Standard errors clustered at the firm level.
Significance levels indicated by: * $p < 0.05$, ** $p < 0.01$, *** $p < 0.001$.

### C.2.3 Different Sectors: Manufacturing vs. services

Tables C.5 and C.6 provide estimation results for quarterly sales growth and quarterly stock returns for the manufacturing and services sectors, respectively, over the periods 2017-2022. The estimated coefficients for sales growth as dependent variable in the manufacturing sector tend to be higher than for the services sector. This result is consistent with the fact that the manufacturing sector is particularly vulnerable to supply chain shocks due to its reliance on complex, global supply chains and just-in-time inventory systems, and supply chains being identified as the most significant issue mentioned by firms on their websites in Figure 5 in the main text.

Table C.5: Regression results corresponding to Eqs. (1) to (3) for log quarterly sales growth and log quarterly stock returns as dependent variables for firms in the manufacturing sector only (NACE 2-digit codes 10 through 33).

|  | Log sales growth (%) | | | Log stock returns (%) | | |
| --- | --- | --- | --- | --- | --- | --- |
|  | (1) | (2) | (3) | (4) | (5) | (6) |
| Covid mention | 0.354 | 0.507 | 0.143 | -0.487* | -0.576** | -0.261 |
|  | (0.519) | (0.517) | (0.546) | (0.202) | (0.202) | (0.206) |
| Affected |  |  |  |  |  |  |
|   Mild | -1.803 | -1.690 | -1.580 | -0.744 | -0.807 | -0.434 |
|  | (1.412) | (1.423) | (1.677) | (0.413) | (0.414) | (0.433) |
|   Moderate | -3.671*** | -3.423** | -3.272* | -1.169** | -1.298*** | -0.795* |
|  | (1.086) | (1.128) | (1.404) | (0.363) | (0.369) | (0.403) |
|   Severe | -3.608* | -3.535* | -3.297 | -0.917 | -1.040* | -0.853 |
|  | (1.505) | (1.523) | (1.869) | (0.497) | (0.500) | (0.551) |
| Lagged log total assets | -6.894*** | -6.737*** | -6.684*** | -6.088*** | -6.167*** | -5.352*** |
|  | (1.007) | (1.010) | (1.095) | (0.273) | (0.273) | (0.251) |
| Workplace closing |  |  |  |  |  |  |
|   Recommended |  | -5.149*** |  |  | 3.441*** |  |
|  |  | (0.727) |  |  | (0.257) |  |
|   Required |  | -6.542*** |  |  | 4.986*** |  |
|  |  | (0.895) |  |  | (0.289) |  |
| Stay at home requirements |  |  |  |  |  |  |
|   Recommended |  | -1.671* |  |  | 2.707*** |  |
|  |  | (0.779) |  |  | (0.249) |  |
|   Required |  | 0.978 |  |  | 1.560*** |  |
|  |  | (0.805) |  |  | (0.275) |  |
| Log fiscal measures |  | 0.194*** |  |  | -0.0800*** |  |
|  |  | (0.0516) |  |  | (0.0140) |  |
| Covid deaths per month |  | 0.0759 |  |  | -0.0589 |  |
|  |  | (0.178) |  |  | (0.0576) |  |
| Firm FE | Yes | Yes | Yes | Yes | Yes | Yes |
| Quarter FE | Yes | Yes | Yes | Yes | Yes | Yes |
| Country-Industry-Quarter FE | No | No | Yes | No | No | Yes |
| No. Firms | 16760 | 16702 | 16416 | 14533 | 14481 | 14178 |
| Observations | 318456 | 317364 | 310489 | 276136 | 275280 | 268075 |
| $R^2$ | 0.0353 | 0.0355 | 0.0675 | 0.133 | 0.136 | 0.292 |

Standard errors in parentheses.
Standard errors clustered at the firm level.
Significance levels indicated by: $^{*}\ p < 0.05$, $^{**}\ p < 0.01$, $^{***}\ p < 0.001$.

Table C.6: Regression results corresponding to Eqs. (1) to (3) for log quarterly sales growth and log quarterly stock returns as dependent variables for firms in the services sector only (NACE 2-digit codes 45 through 96).

|  | Log sales growth (%) | | | Log stock returns (%) | | |
|---|---|---|---|---|---|---|
|  | (1) | (2) | (3) | (4) | (5) | (6) |
| Covid mention | 0.304 | 0.365 | 1.392 | -0.424 | -0.612* | -0.616* |
|  | (0.811) | (0.815) | (0.889) | (0.259) | (0.259) | (0.271) |
| Affected |  |  |  |  |  |  |
|   Mild | 0.217 | 0.266 | 0.860 | -0.569 | -0.616* | -0.957** |
|  | (0.804) | (0.817) | (1.001) | (0.290) | (0.291) | (0.308) |
|   Moderate | -1.988** | -2.003** | -0.920 | -0.738* | -0.769* | -0.974** |
|  | (0.705) | (0.726) | (0.927) | (0.309) | (0.311) | (0.338) |
|   Severe | -3.207** | -3.041** | -3.115** | -0.999* | -1.119** | -1.392** |
|  | (1.047) | (1.067) | (1.194) | (0.403) | (0.404) | (0.441) |
| Lagged log total assets | -5.913*** | -5.920*** | -5.088*** | -3.491*** | -3.419*** | -3.924*** |
|  | (0.964) | (0.966) | (1.083) | (0.688) | (0.692) | (0.274) |
| Workplace closing |  |  |  |  |  |  |
|   Recommended |  | -3.943*** |  |  | 4.966*** |  |
|  |  | (0.837) |  |  | (0.265) |  |
|   Required |  | -6.509*** |  |  | 7.624*** |  |
|  |  | (0.981) |  |  | (0.314) |  |
| Stay at home requirements |  |  |  |  |  |  |
|   Recommended |  | -3.954*** |  |  | 2.383*** |  |
|  |  | (0.699) |  |  | (0.233) |  |
|   Required |  | -3.956*** |  |  | -0.472 |  |
|  |  | (0.881) |  |  | (0.300) |  |
| Log fiscal measures |  | 0.0857 |  |  | 0.0323* |  |
|  |  | (0.0442) |  |  | (0.0139) |  |
| Covid deaths per month |  | -0.0259 |  |  | -0.0777 |  |
|  |  | (0.134) |  |  | (0.0490) |  |
| Firm FE | Yes | Yes | Yes | Yes | Yes | Yes |
| Quarter FE | Yes | Yes | Yes | Yes | Yes | Yes |
| Country-Industry-Quarter FE | No | No | Yes | No | No | Yes |
| No. Firms | 12915 | 12857 | 12301 | 15463 | 15381 | 14808 |
| Observations | 240979 | 240017 | 227442 | 294853 | 293482 | 280433 |
| $R^2$ | 0.0484 | 0.0490 | 0.143 | 0.109 | 0.112 | 0.305 |

Standard errors in parentheses.
Standard errors clustered at the firm level.
Significance levels indicated by: * $p < 0.05$, ** $p < 0.01$, *** $p < 0.001$.

### C.2.4 Lagged Dependent Variable

Table C.7 provides estimation results for quarterly sales growth in columns (1) to (3) and quarterly stock returns in columns (4) to (6) as dependent variable including the lagged dependent variable as an additional covariate over the periods 2017-2022. Comparing the estimation results in Table C.7 with the ones in Table C.1 shows that the estimates are robust and there is little concern due to endogeneity from persistent time trends. Moreover, estimating fixed effects in dynamic panel data in the "small $T$, large $N$" context with a lagged dependent variable as covariate can introduce a "Nickell bias".[34] Since $T$ is 24 periods in the main estimates, the upper limits for this possible bias are small, Nickell shows that they are approximately $-(1 + \beta_{\text{lagged dep.}})/(T-1)$ which in our case equals roughly $-0.06$ for sales growth and $-0.05$ for stock returns.

**Table C.7**: Regression results with one lag of the dependent variable.

|  | Log sales growth (%) | | | Log stock returns (%) | | |
|---|---|---|---|---|---|---|
|  | (1) | (2) | (3) | (4) | (5) | (6) |
| Lagged dependent | -0.306*** | -0.306*** | -0.310*** | -0.106*** | -0.106*** | -0.105*** |
|  | (0.00783) | (0.00786) | (0.00849) | (0.00482) | (0.00486) | (0.00503) |
| Covid mention | 0.551 | 0.736 | 0.811 | -0.527*** | -0.664*** | -0.489** |
|  | (0.433) | (0.436) | (0.471) | (0.160) | (0.160) | (0.165) |
| Affected |  |  |  |  |  |  |
|   Mild | -1.240 | -1.172 | -0.347 | -0.716** | -0.810*** | -0.747** |
|  | (0.728) | (0.738) | (0.936) | (0.231) | (0.232) | (0.247) |
|   Moderate | -3.687*** | -3.589*** | -2.492** | -1.120*** | -1.230*** | -0.897*** |
|  | (0.587) | (0.609) | (0.812) | (0.229) | (0.232) | (0.257) |
|   Severe | -4.769*** | -4.684*** | -3.486*** | -1.444*** | -1.622*** | -1.292*** |
|  | (0.797) | (0.810) | (1.012) | (0.303) | (0.304) | (0.337) |
| Lagged log total assets | -4.359*** | -4.258*** | -3.929*** | -4.551*** | -4.531*** | -4.584*** |
|  | (0.666) | (0.666) | (0.738) | (0.453) | (0.455) | (0.205) |
| Workplace closing |  |  |  |  |  |  |
|   Recommended |  | -5.067*** |  |  | 4.102*** |  |
|  |  | (0.523) |  |  | (0.182) |  |
|   Required |  | -6.767*** |  |  | 6.087*** |  |
|  |  | (0.604) |  |  | (0.210) |  |
| Stay at home requirements |  |  |  |  |  |  |
|   Recommended |  | -3.240*** |  |  | 2.464*** |  |
|  |  | (0.476) |  |  | (0.165) |  |
|   Required |  | -2.157*** |  |  | 0.423* |  |
|  |  | (0.539) |  |  | (0.200) |  |
| Log fiscal measures |  | 0.121*** |  |  | 0.00423 |  |
|  |  | (0.0296) |  |  | (0.00947) |  |
| Covid deaths per month |  | 0.0816 |  |  | -0.0927** |  |
|  |  | (0.0983) |  |  | (0.0353) |  |
| Firm FE | Yes | Yes | Yes | Yes | Yes | Yes |
| Quarter FE | Yes | Yes | Yes | Yes | Yes | Yes |
| Country-Industry-Quarter FE | No | No | Yes | No | No | Yes |
| No. Firms | 32344 | 32214 | 30356 | 32749 | 32599 | 31009 |
| Observations | 604794 | 602564 | 565409 | 620058 | 617651 | 584005 |
| $R^2$ | 0.132 | 0.132 | 0.187 | 0.129 | 0.132 | 0.310 |

Standard errors in parentheses
Standard errors clustered at the firm level
* $p < 0.05$, ** $p < 0.01$, *** $p < 0.001$